%% file: PRD_follow_up-060120.tex
\documentclass[aps,twocolumn,prd,amsmath,nofootinbib,superscriptaddress]{revtex4-1}

\usepackage{epsfig,slashed,nicefrac}
\usepackage[utf8]{inputenc}
\usepackage{color}
\usepackage{multirow}
\usepackage{lineno}
\usepackage{float} 
\usepackage[caption=false]{subfig} 
\usepackage{eqnarray,amsmath}
\usepackage{paralist}
\usepackage{times}
\usepackage{txfonts}
\usepackage{comment}
\usepackage{xspace}
\usepackage[normalem]{ulem}

\usepackage{scalerel}
\usepackage{tikz}
\usetikzlibrary{svg.path}

\definecolor{orcidlogocol}{HTML}{A6CE39}
\tikzset{
  orcidlogo/.pic={
    \fill[orcidlogocol] svg{M256,128c0,70.7-57.3,128-128,128C57.3,256,0,198.7,0,128C0,57.3,57.3,0,128,0C198.7,0,256,57.3,256,128z};
    \fill[white] svg{M86.3,186.2H70.9V79.1h15.4v48.4V186.2z}
                 svg{M108.9,79.1h41.6c39.6,0,57,28.3,57,53.6c0,27.5-21.5,53.6-56.8,53.6h-41.8V79.1z M124.3,172.4h24.5c34.9,0,42.9-26.5,42.9-39.7c0-21.5-13.7-39.7-43.7-39.7h-23.7V172.4z}
                 svg{M88.7,56.8c0,5.5-4.5,10.1-10.1,10.1c-5.6,0-10.1-4.6-10.1-10.1c0-5.6,4.5-10.1,10.1-10.1C84.2,46.7,88.7,51.3,88.7,56.8z};
  }
}

\newcommand\orcidicon[1]{\href{https://orcid.org/#1}{\mbox{\scalerel*{
\begin{tikzpicture}[yscale=-1,transform shape]
\pic{orcidlogo};
\end{tikzpicture}
}{|}}}}

\newcommand{\eg}{{\it e.g.}}
\newcommand{\ie}{{\it i.e.}}
\newcommand{\cf}[1]{{Fig.~\ref{#1}}}
\def\sqrtsNN {\mbox{$\sqrt{s_{NN}}$}\xspace}

\def\RpA     {\mbox{$R_{pA}$}}

\def\RdAu    {\mbox{$R_{d\rm Au}$}}
\def\RpAu    {\mbox{$R_{p\rm Au}$}}

\def\RpPb    {\mbox{$R_{p\rm Pb}$}}

\def\pPb  {$p\mathrm{Pb}$}
\def\ycms  {y_{\rm c.m.s.}}

\def\be{\begin{equation*}}
\def\ee{\end{equation*}}
\def\bsp#1\esp{\begin{split}#1\end{split}} 
\def\bpm{\begin{pmatrix}}
\def\epm{\end{pmatrix}}

 \usepackage{ifpdf}
 \ifpdf
 \usepackage[pdftex]{hyperref}
 \else
 \usepackage[hypertex]{hyperref}
 \fi

 \hypersetup{
   pdftitle={},%
   pdfauthor={},%
   pdfsubject={},%
   pdfkeywords={},%
   pdfstartview={},%
   bookmarksopen=true, breaklinks=true, debug=true, %
   colorlinks=true, linkcolor=red, citecolor=blue, urlcolor=blue
 }

\begin{document}

\title{Reweighted nuclear PDFs using Heavy-Flavor Production Data at the LHC: \\ $\mathbf{nCTEQ15_{rwHF}}$ \& $\mathbf{ EPPS16_{rwHF}}$}

\author{Aleksander Kusina\orcidicon{0000-0002-4090-0084}}
\affiliation{Institute of Nuclear Physics, Polish Academy of Sciences, ul. Radzikowskiego 152, 31-342 Cracow, Poland}

\author{Jean-Philippe Lansberg\orcidicon{0000-0003-2746-5986}}
\affiliation{Universit\'e Paris-Saclay, CNRS, IJCLab, 91405 Orsay, France}

\author{Ingo Schienbein\orcidicon{0000-0003-0373-474X}}
\affiliation{Laboratoire de Physique Subatomique et de Cosmologie,
              Universit\'e Grenoble-Alpes, CNRS/IN2P3,
              53 avenue des Martyrs, 38026 Grenoble, France}

\author{Hua-Sheng Shao\orcidicon{0000-0002-4158-0668}}
\affiliation{Laboratoire de Physique Th\'eorique et Hautes Energies (LPTHE), UMR 7589, Sorbonne Universit\'e et CNRS, 4 place Jussieu, 75252 Paris, France}

\date{\today}

\begin{abstract}
We present the reweighting of two sets of nuclear PDFs, nCTEQ15 and EPPS16, using a selection of experimental data on heavy-flavor meson
[$D^0$, $J/\psi$, $B\rightarrow J/\psi$ and $\Upsilon(1S)$] production in proton-lead collisions at the LHC which were not used in the original determination of these nuclear PDFs. The reweighted PDFs exhibit significantly smaller uncertainties thanks to these new heavy-flavor  constraints. We present a comparison with another selection of data from the LHC and RHIC which were not included in our reweighting procedure. The comparison is overall very good and serves as a validation of these reweighted nuclear PDF sets,which we dub $\rm nCTEQ15_{rwHF}$ \& $\rm EPPS16_{rwHF}$. This indicates that the LHC and forward RHIC heavy-flavor data can be described within the standard collinear factorization framework with the same (universal) small-$x$ gluon distribution. We discuss how we believe such reweighted PDFs should be used as well as the limitations of our procedure.
\end{abstract}

\maketitle

\section{Introduction}
\input{introduction-171220.tex}

\section{Context and methodology}
\input{context-methodology-171220.tex}

\section{Resulting reweighted \lowercase{n}PDFs}
\input{reweighted-nPDF-171220.tex}

\section{Results}
\label{sec:res}
\input{predictions-171220.tex}

\section{Conclusions \label{sec:conclusion}}
\input{conclusions-171220.tex}


{\bf Acknowledgments.} We are grateful to R.~Abdul Khalek, N.~Armesto, Y.~Zhang for very helpful discussions. 
This project has received funding from the European Union’s Horizon 2020 research and innovation programme under the grant agreement No.824093 in order to contribute to the EU Virtual AccessNLOAccess. This work was also partly supported by the French CNRS via the IN2P3 project GLUE@NLO, via the IEA Excitonium, by the Paris-Saclay U. via the P2I Department and by the P2IO Labex via the Gluodynamics project. A.K. is also grateful for the support of Narodowe Centrum Nauki under grant SONATAbis no 2019/34/E/ST2/00186.

\bibliographystyle{utphys}
\bibliography{PRD_follow_up-171220}

\vspace*{-.5cm}
\appendix
\input{LHAPDFfiles-211220.tex}

\end{document}

%% file: introduction-171220.tex
In~\cite{Kusina:2017gkz}, we investigated how experimental data on the inclusive production of heavy-flavor (HF) mesons
[$D^0$, $J/\psi$, $B\rightarrow J/\psi$ and $\Upsilon(1S)$]
in proton-lead collisions at the LHC could advance our knowledge 
of the gluon-longitudinal-momentum distribution inside heavy nuclei.
Indeed, at the LHC and RHIC energies, such HF reactions are usually initiated by gluons
and their production in proton-nucleus collisions should shed some light
on the nuclear gluon content. \footnote{Recently, associated HF reactions were also shown~\cite{Shao:2020acd,Shao:2020kgj} to be good probes of the impact-parameter-dependent partonic nuclear content.}

We observed that the nuclear effects encoded in two recent global fits of 
nuclear parton densities at next-to-leading order (NLO), nCTEQ15~\cite{Kovarik:2015cma} and EPPS16~\cite{Eskola:2016oht}, were yielding
a good description of the existing LHC HF production data
supporting the hypothesis that the modification of the partonic densities in heavy nuclei
could be the dominant effect at play in such processes.

We then went further in our investigations and performed a Bayesian-reweighting analysis 
of the data sample of each of the mesons and showed that the existing HF data were clearly pointing
at a depleted gluon distribution at small momentum fractions, $x$, in the lead nucleus, also
known as shadowing. According to our reweighting analysis, the
significance of this depletion was larger than 7 $\sigma$ at $x$ smaller than 0.01.
In addition, our analysis also supported the existence of gluon antishadowing, 
whereby the gluon content of nuclei is augmented when $x \simeq 0.1$.

We concluded that the inclusion of such HF data in a nuclear PDF (nPDF) fit, such as those used for our analysis, would reduce the uncertainty on the gluon density in heavy nuclei down to  $x\simeq 7\times 10^{-6}$. At such low $x$, there is currently simply no other data. We stressed that the reweighted nPDFs would still be compatible with the other data of the global fits, in particular in the quark sector. 

Yet, we noted that the large factorization scale ($\mu_F$) uncertainty in the computation of HF-production cross sections and nuclear modification factors (NMFs) was then to be considered. We indeed found it to generate an uncertainty larger than that from reweighted nPDFs using charm or charmonium data. Indeed, the magnitude of the nuclear effects encoded in the nPDFs depend on the scale via the Dokshitzer-Gribov-Lipatov-Altarelli-Parisi (DGLAP) evolution equation and such an evolution is admittedly fast in the GeV range.

In the present work, we follow up on our previous study published as a Letter. First, we provide more details about the reweighting analysis and new comparisons with data sets which became available after our study was performed. They serve as post validation of our reweighting analysis.  Second we have converted our reweighted nPDFs, which were initially in the form of replicas along with weights, into nPDFs with Hessian uncertainties. As such, they can be presented in files to be included and easily used in the LHAPDF library. A new feature of our analysis is that it includes the effect of the factorization-scale choice. As such, we also provide specific guidances on how to use them for future studies, in particular to predict the initial-state nuclear effect on hard probes of the quark-gluon plasma (QGP).

%% file: context-methodology-171220.tex
\subsection{Framework and its justification}

In the collinear factorization, following Feynman's model of partons, the longitudinal-momentum distributions of quarks and gluons inside hadrons are given by the PDFs. These then connect the measurable hadronic cross sections and the partonic cross sections (induced by quarks and gluons) which can be calculated using perturbative methods. 

The determination of PDFs of free protons, denoted $f_i^p$, is actively pursued by many colleagues via global analyses of as much as possible experimental data of hard processes which are believed to be perturbative enough. Such state-of-the-art global analyses~\cite{Hou:2019efy,Harland-Lang:2014zoa,Ball:2017nwa,Accardi:2016qay,Alekhin:2017kpj,Alekhin:2014irh} involve complex perturbative calculations and cutting-edge statistical methods to extract PDFs and their uncertainties altogether. 

Assuming the same collinear factorization to apply in proton-nucleus ($pA$) and nucleus-nucleus ($AA$) collisions, one needs to introduce nPDFs, denoted $f_i^{A}$. 
In such a case, additional challenges come out along with additional motivations to study them. Indeed, nuclear data are significantly more complex to collect and one needs to cope with two additional degrees of freedom to describe the nuclei, the number of protons ($Z$) and neutrons ($N=A-Z$)  which they comprise. These are also necessary inputs to employ hard probes of the QGP produced in ultra-relativistic heavy-ion collisions at RHIC and the LHC~\cite{Andronic:2015wma}. Constraining nPDFs is then not only about the ambitious endeavor to understand the quark and gluon content of the nuclei but also to understand the initial stages of the production of some QGP probes.

Since the early 1980's, we know that the partonic description of nuclei cannot be reduced to a simple collection of
partons in free nucleons. In other words, nPDFs deviate from a simple sum of nucleon PDFs.
To study such deviations, it is customary to rely on NMFs, like 
\be
R[F_2^{\ell A}]= \frac{F_2^{\ell A}}{Z F_2^{\ell p}+N F_2^{\ell n}}
\ee
for the DIS structure function $F_2$
and parton-level NMFs 
\be
R_i^A(x,\mu_F) = \frac{f_i^A}{Z f_i^p + N f_i^n}
\ee 
with 
$f_i^A \equiv Z f_i^{p/A} + N f_i^{n/A}$, instead of the absolute nPDFs.

Past studies of $F_2$ \cite{Aubert:1983xm,Goodman:1981hc,Bodek:1983ec,Bari:1985ga,Benvenuti:1987az,Ashman:1988bf,Arneodo:1988aa}
told us that, for the {\it quarks},
\begin{enumerate}[(i)]\setlength{\itemsep}{-0.1cm}
\item $R_q^A>1$ for $x\gtrsim0.8$ (Fermi-motion region), 
\item $R_q^A<1$ for $0.25\lesssim x\lesssim0.8$ (EMC region), 
\item $R_q^A>1$ for $0.1\lesssim x\lesssim0.25$ (antishadowing region), and 
\item $R_q^A<1$ for $x\lesssim0.1$ (shadowing region).
\end{enumerate}
These 4 different $x$ regions are usually denoted by the names in parenthesis.  The EMC region still lacks a fully conclusive picture~\cite{Higinbotham:2013hta} although the region of medium and large $x$, $R_q^A$ is usually explained by nuclear-binding and medium effects and the Fermi motion of the nucleons \citep{Geesaman:1995yd}. At small $x$, coherent scatterings inside the nucleus 
 explain the observed suppression of $F_2$. This is why it is referred to as shadowing. The physics underlying the antishadowing is however less firmly established. 

Although the generic trend of the nPDFs could in principle be related to specific physics phenomena, they  remain essentially determined by global fits of experimental data \cite{Eskola:2016oht,Kovarik:2015cma,Khanpour:2016pph,deFlorian:2011fp,Hirai:2007sx,AbdulKhalek:2020yuc,Walt:2019slu,Kusina:2020lyz} based on initial parameterizations only slightly driven by the above physics considerations.

 The above discussion specifically relates to the nuclear {\it quark} content  which can directly be probed by lepton-nucleus ($\ell A$) DIS and $pA$ Drell-Yan processes. Lacking corresponding direct probes, the nuclear {\it gluon} content is less known, despite indirect constraints from the scaling violation~\cite{Kovarik:2015cma}.

As such, the NLO nPDF fits, nCTEQ15~\cite{Kovarik:2015cma} and EPPS16~\cite{Eskola:2016oht},
were performed using data from RHIC on single inclusive pion production. In the case of EPPS16, jet data from the LHC were also used. The objective was precisely  to constrain the gluon densities down to $x\sim 10^{-3}$. However, owing to the absence of data at $x \lesssim 10^{-3}$,
the nuclear gluon content remained completely undetermined at small $x$. As such, the resulting gluon nPDFs in this region are mere extrapolations from the region of larger $x$ and essentially follow from  their $x$-dependent parameterizations at the scale $\mu_{F,0} \sim 1$~GeV where the perturbative DGLAP evolution is initiated. 

As pointed out several times~\cite{Stavreva:2010mw,Helenius:2016hcu}, the uncertainties derived from these gluon 
nPDFs is not representative of the true uncertainty on this quantity. More flexible initial nPDF parameterizations naturally yield
 much larger uncertainties in this region. This explains why the EPPS16 set, despite accounting for more data constraints, 
show much larger uncertainties with respect to its predecessor EPS09~\cite{Eskola:2009uj} or nCTEQ15.

In this context and motivated by the results of proton studies using HF production to improve the determination of small-$x$ gluon PDF~\cite{Zenaiev:2015rfa,Gauld:2015yia,Cacciari:2015fta,Gauld:2016kpd,deOliveira:2017ega}, we thus studied~\cite{Kusina:2017gkz} the impact of HF production to constrain the small-$x$ gluon density in lead down to $x \simeq 7 \times 10^{-6}$. In particular, we used  heavy-quark and heavy-quarkonium data in LHC proton-lead ($p$Pb) collisions. 

Like the follow-up work presented here, our former study relied on the assumption that collinear factorization in terms of nPDFs holds in the nuclear environment. To date, the global usage of nPDFs is still a subject of debates. Such an assumption should thus be seen as a working hypothesis
which has to be systematically questioned. This is the object of this extension with new data-theory comparisons and
the release of reweighted nPDFs which can serve for future studies which will confirm or falsify this framework.

Indeed, we recall that other Cold-Nuclear Matter (CNM) effects~\cite{Gerschel:1988wn,Vogt:1999cu,Ferreiro:2014bia,Ferreiro:2018wbd,Capella:2005cn,Capella:2000zp,Gavin:1990gm,Brodsky:1989ex,Ducloue:2015gfa,Ma:2015sia,Fujii:2013gxa,Qiu:2013qka,Kopeliovich:2001ee,Ferreiro:2008wc,Ferreiro:2011xy,Ferreiro:2013pua,Vogt:2010aa}
could be at play in specific conditions, in particular for the quarkonium case.
In the adopted framework, such additional effects are however considered  as Higher-Twist (HT) contributions and our
working assumption can be seen as the consideration of the sole of the Leading-Twist (LT) factorizable contributions.

\subsection{Connecting NMFs and \lowercase{n}PDFs}

Since the advent of RHIC and the LHC, thus for two decades now, the cross-section measurements performed at $pA$ colliders 
have nearly systematically be normalized to the $pp$ ones~\cite{Albacete:2016veq,Andronic:2015wma,Albacete:2017qng}.
Indeed,  the prime interest of such studies is to look for deviations from the free nucleon case, up to isospin effects. 

Hard-perturbative reactions are rare and each of the $\langle N_{\rm coll.}\rangle$ binary nucleon-nucleon ($NN$) collisions triggered by a $pA$ or $AA$ collision is meant to independently and equally contribute to the observed yields $Y$: $Y_{pA / AA} \simeq  \langle N_{\rm coll.}\rangle Y_{NN}$. If one considers all the possible geometrical configurations for  these collisions, this expected equality can be translated 
into a relation between the cross sections, namely ${d\sigma_{pA}} \simeq{A \times d\sigma_{pp}}$. As such, it is natural
to define the NMF
\be
\RpA\equiv\frac{d\sigma_{pA}}
{A \times d\sigma_{pp}},
\ee
such that (up to isospin effects) it would equate unity in the absence of nuclear effects. Just as $R_q^A$ was defined above, one can define $R_g^A$ with the difference that gluon densities are a priori identical in protons and neutrons. If this was the only nuclear effect at play, $\RpA$ would then directly be connected to $R_g^A$ via the integration of kinematic variables. Only for specific reactions at leading order (LO) can one write an equality. In general, it is necessarily more complex.

In the case of DIS off a nucleus $A$, one historically used $F_2^{\ell A}$ instead of $\sigma$ and the corresponding NMF $R[F_2]$ is then naturally found to probe the modification of the (anti)quark nPDF compared to its PDF, \ie\ $R_q^A$ as we discussed above.

The NMFs at the LHC and RHIC are so far differential in the transverse momentum ($P_{T,\mathcal{H}}$) or the center-of-momentum (cms)
rapidity $y_{cms,\mathcal{H}}$ of the observed hadron $\mathcal{H}$ as well as a function of the collision centrality. The latter remains theoretically poorly understood and introduces many complications. As such, it was not considered in~\cite{Kusina:2017gkz}. Since the current study  bears on this first study, we leave the centrality dependence for future investigations and focus on centrality integrated results where the geometry of the collisions should not matter.

Focusing on $\RpA$ is justified by the following. First, it essentially removes the sensitivity
from the theoretical side on the {\it proton} PDF. Indeed, at very small $x$, the PDF
uncertainties are not necessarily negligible. Second, $\RpA$ is a priori less sensitive
to the modification of the normalization of the 
cross-section by QCD corrections to the hard parton scattering. Third, on the experimental side, 
$\RpA$ is usually better determined than the  $pA$ yields because 
some systematic experimental uncertainties can be assumed to cancel.

As we have already alluded to, the connection between $\RpA$ and $R_{q,g}^A$ is not necessarily trivial even assuming that
the reaction is only initiated by gluon fusion as expected at high energy for HF production.
Since we follow the procedure of~\cite{Kusina:2017gkz}, we will employ the data-driven approach of~\cite{Lansberg:2016deg,Shao:2012iz,Shao:2015vga} where the matrix  elements squared $|A|^2$ for the gluon-fusion processes 
are determined from $pp$ data restricting to a $2\to2$ kinematics.\footnote{See~\cite{Ferreiro:2008wc} for a discussion of the relevance
of a $2\to2$ kinematics to predict $J/\psi$ $\RpA$ as a function of $y$.} 
First and foremost, it is justified by our limited understanding of the quarkonium-production mechanisms (see \eg\ \cite{Lansberg:2019adr,Andronic:2015wma,Brambilla:2010cs}) while being sufficient to perform a sound evaluation of the nPDF effects via $\RpA$. Second, the same approach also applies to open HF hadrons~\cite{Lansberg:2016deg} for which full-fledged perturbative QCD computations exist\footnote{GM-VFNS~\cite{Kniehl:2004fy,Kniehl:2005mk,Kniehl:2012ti,Kniehl:2011bk,Kniehl:2015fla}, MG5aMC~\cite{Alwall:2014hca} and FONLL~\cite{Cacciari:1998it,Cacciari:2001td,Cacciari:2012ny}.}. Indeed, we have checked that FONLL yields equivalent results. In practice, it boils down to employ a specific empirical functional form for $|A|^2$, initially used in~\cite{Kom:2011bd} to model single-quarkonium hadroproduction for double parton scattering studies~\cite{Kom:2011bd,Lansberg:2014swa,Lansberg:2015lva,Shao:2016wor,Borschensky:2016nkv,Shao:2019qob,Shao:2020kgj}. The latter is flexible enough to provide a fair account of HF data on single-inclusive-particle production. In fact, this is not always the case for complete perturbative QCD (pQCD) computations which have however the virtue of being derived from first principles of QCD.

The case of $D$ meson production in $p$Pb collisions has in fact recently been studied in detail~\cite{Eskola:2019bgf}  using the SACOT-$m_T$ variant~\cite{Helenius:2018uul} of the GM-VFNS formalism. It has been found to yield the same qualitative features as what we obtained in~\cite{Kusina:2017gkz} regarding the constraints on the nCTEQ15 and EPPS16 nPDFs and the rather large
associated factorisation-scale uncertainty.

The employed data-driven approach is also such that
\begin{enumerate}\setlength{\itemsep}{-0.1cm}
\item the event generation is much faster than using NLO QCD-based codes. This allows us to perform computations 
for several nPDFs (2 in our case) with 3 scale choices for 4 particles in an acceptable amount of computing time.
\item one can employ it directly for any single-inclusive-particle spectrum once we know
the relative contribution of different $ij$ fusion channels, \ie\ the parton luminosities times $|A_{ij}|^2$. 
Thus, in principle, quark channels can also be accounted for. 
\item  the normalization uncertainty in the $pp$ cross section is controlled by the measured data, which also enters \RpA.
\end{enumerate}

Significant theoretical uncertainties arise from the $\mu_F$ uncertainty in these QCD processes. Different reweighted nPDF sets therefore naturally emerge from the scale variation and we will explain how they should be used in section~\ref{sec:res}.

We stick to the same $pp$ baseline as in~\cite{Kusina:2017gkz} which was slightly extended and improved compared to~\cite{Lansberg:2016deg}. It includes the non-prompt $J/\psi$ from $B \to J/\psi$ data. For the $D^0$, $J/\psi$ and $\Upsilon(1S)$, a scale variation in the $pp$ baseline itself was performed, besides that in $R_g^{\text{Pb}}(x,\mu_F)$. We also had checked that for $D^0$ and  $B\rightarrow J/\psi$ production, the scale uncertainty on \RpPb\ nearly matches that of FONLL whereas FONLL shows significantly larger scale uncertainties on the {\it cross section}. This discussion was given as supplemental material to~\cite{Kusina:2017gkz}.\footnote{See section~B and Fig.~1 of \url{https://journals.aps.org/prl/supplemental/10.1103/PhysRevLett.121.052004/appendix.pdf}.}

\subsection{The PDF reweighting method}

PDF reweighting provides the means for estimating the impact of new experimental data on PDFs without performing a full scale global analysis. Instead, Bayes theorem is adopted to update the underlying probability distribution and obtain a new updated set of PDFs which takes into account information from the new data. The first use of the reweighting in the context of a PDF analysis was concluded in~\cite{Giele:1998gw}. Further development of this technique for PDF analyses was carried out resulting in a number of improvements/variants of this method~\cite{Ball:2010gb,Ball:2011gg,Sato:2013ika,Paukkunen:2014zia}.

For the purpose of the study in~\cite{Kusina:2017gkz} and the current analysis we used the reweighting procedure outline in~\cite{Kusina:2016fxy} which we briefly summarize here. In the first step, a Hessian PDF set is converted into PDF replicas which have a direct probabilistic interpretation. Since the nPDFs we are using (nCTEQ15 and EPPS16) provide symmetric errors, the conversion allows for an arbitrarily precise reproduction of the corresponding uncertainties\footnote{The precision depends only on the number of used replicas which, in our case, is $10^4$.} and it can be obtained using the following formula
\begin{equation}
  f_k = f_0 + \sum_{i=1}^{N} \frac{f_i^{(+)}-f_i^{(-)}}{2} R_{ki},
\end{equation}
where $f_k$ are\footnote{To be precise, $f$ stands for a vector of functions in the flavor space.} PDF replicas, $f_0$ is the central PDF from the Hessian set,
$f_i^{(+)/(-)}$ are the corresponding Hessian error PDFs, $N$ is the number of
eigen-vector directions for the Hessian set, and $R_{ki}$ are normally-distributed
random numbers centered at 0 with a standard deviation of 1.
For such a set of replicas, one can compute the average (corresponding to the
Hessian central value) and variance of a PDF-dependent observable ${\cal O}$ as
\begin{equation}
  \begin{split}
    \left<{\cal O}\right> & = \frac{1}{N_{\text{rep}}}\sum_{k=1}^{N_{\text{rep}}}{\cal O}(f_{k}),\\
    \delta\left<{\cal O}\right> & = \sqrt{\frac{1}{N_{\text{rep}}}\sum_{k=1}^{N_{\text{rep}}}
      \left({\cal O}(f_{k})-\left<{\cal O}\right>\right)^{2}},
  \end{split}
\end{equation}
where $N_{\mathrm{rep}}$ is the number of replicas.

During the reweighting, each of the PDF replicas is supplemented with a corresponding
weight which depends on how well the replica describes the new data. In our analysis,
the weight is defined as
\begin{equation}
w_{k}=\frac{e^{-\frac{1}{2}\chi_{k}^{2}/T}}
           {\frac{1}{N_{\text{rep}}}\sum_{i}^{N_{\text{rep}}}e^{-\frac{1}{2}\chi_{i}^{2}/T}},
\label{eq:weight}
\end{equation}
where $T$ is the tolerance criterion used when defining Hessian
error PDFs~\cite{Pumplin:2001ct}
and $\chi_{i}^{2}$ quantifies how well the new data is described.
After calculating the weights, it is straightforward to compute any PDF-dependent
observables as
\begin{equation}
  \begin{split}
    \left<{\cal O}\right>_{\text{new}} & =
          \frac{1}{N_{\text{rep}}}\sum_{k=1}^{N_{\text{rep}}}w_{k}{\cal O}(f_{k}),\\
    \delta\left<{\cal O}\right>_{\text{new}} & =
          \sqrt{\frac{1}{N_{\text{rep}}}\sum_{k=1}^{N_{\text{rep}}}w_{k}
          \left({\cal O}(f_{k})-\left<{\cal O}\right>_{\text{new}}\right)^{2}}.
\end{split}
\label{eq:ave-rew}
\end{equation}
More details on the reweighting procedure can be found in~\cite{Kusina:2016fxy}
and in the other aforementioned references.

\subsection{Data selection} 

In a global PDF fit HT corrections have to be included in the theory predictions or kinematic cuts have to be imposed such that HT effects are expected to be small in the selected data. For our study, one also needs to select a kinematical region where gluon fusion is the dominant channel and where nPDFs represent the main nuclear effects. This is why we decided in~\cite{Kusina:2017gkz} to focus on open- and hidden-HF production in $pA$ collisions at LHC energies. Due to the large Lorentz boost at these energies, as what regards quarkonium production, the heavy-quark pair remains small while traversing the nuclear matter. As such, the break-up of the pair~\cite{Vogt:2004dh,Lourenco:2008sk} is negligible at the LHC whereas it is a potentially large effect at low(er) energies. Along the same lines, we have not considered the more fragile excited states ($\psi(2S)$, $\Upsilon(2S)$, $\Upsilon(3S)$) and instead we focused on $J/\psi$ and $\Upsilon(1S)$ data on which the comover effects~\cite{Ferreiro:2018wbd,Ferreiro:2014bia,Capella:2005cn,Capella:2000zp,Gavin:1990gm} are likely limited.

Overall, we have used the following datasets: the  ALICE~\cite{Abelev:2014hha} and LHCb~\cite{Aaij:2017gcy} $D^0$ data;
the  ALICE~\cite{Adam:2015iga,Abelev:2013yxa} and LHCb~\cite{Aaij:2013zxa,Aaij:2017cqq} $J/\psi$ data;
the LHCb~\cite{Aaij:2017cqq} $B\rightarrow J/\psi$ data; the 
 ALICE~\cite{Abelev:2014oea}, ATLAS~\cite{TheATLAScollaboration:2015zdl}  and LHCb~\cite{Aaij:2014mza} $\Upsilon(1S)$ data.
The $\chi^2$ we obtained and the original values are given as supplemental material to~\cite{Kusina:2017gkz}.\footnote{They can be found in Table 2 
of \url{https://journals.aps.org/prl/supplemental/10.1103/PhysRevLett.121.052004/appendix.pdf}.}

Forward $d$Au $J/\psi$ RHIC data could have been added. Instead, we preferred to focus
on the LHC data at 5 and 8 TeV and to use the RHIC~\cite{Adare:2010fn,Adare:2012qf} and the new LHC~\cite{Aaboud:2017cif,Adam:2016ich} ones as a cross check as we do here.

%% file: reweighted-nPDF-171220.tex
In~\cite{Kusina:2017gkz}, we have performed a PDF reweighting using
two sets of initial nPDFs: nCTEQ15 and EPPS16, and for each of these nPDFs
we have used data for $J/\psi$, $D$-meson, $B\to J/\psi$ and $\Upsilon(1S)$
production from $p$Pb collisions at the LHC to produce four reweighted nPDF (RnPDF) sets. 
The reweighting for each data set
was done independently and 3 sets were obtained from a scale
variation using the following scale choices: $\mu=\{\mu_0, 2\mu_0, 0.5\mu_0\}$, see Tab.~\ref{tab:scale}
for the values of $\mu_0$.

Going further than in~\cite{Kusina:2017gkz}, we have created, for the current study, an additional RnPDF set 
for each set of data by combining the uncertainties obtained from the scale variations. The combined set is produced by simply taking all replicas from reweightings with different scale choices and treating them as equally
probable.
\begin{table}[hbt!]
\begin{tabular}{ccccc}
\hline\hline
          & $J/\psi$ & $D$ & $B\to J/\psi$ & $\Upsilon(1S)$ \\ 
\hline
$\mu_0^2$ & $M_{J/\psi}^2+P_{T,{J/\psi}}^2$ & $4 M_{D}^2+P_{T,{D}}^2$ & $4M_B^2 + \frac{M_B^2}{M_{J/\psi}^2} P_{T,J/\psi}^2$  & $M_{\Upsilon}^2+P_{T,{\Upsilon}}^2$ \\
\hline\hline
\end{tabular}
\caption{Central scale choice squared, $\mu^2_0$, for the considered processes.
Note that for scale variation we use $\mu_F = \xi\mu_0$ with
$\xi = \{1,2,0.5\}$.}
\label{tab:scale}
\end{table}

\begin{figure}[!htb]
\centering{}
\subfloat[$D$-reweighted nCTEQ15\label{fig:nCTEQ15compDscale}]{
\includegraphics[width=0.5\columnwidth]{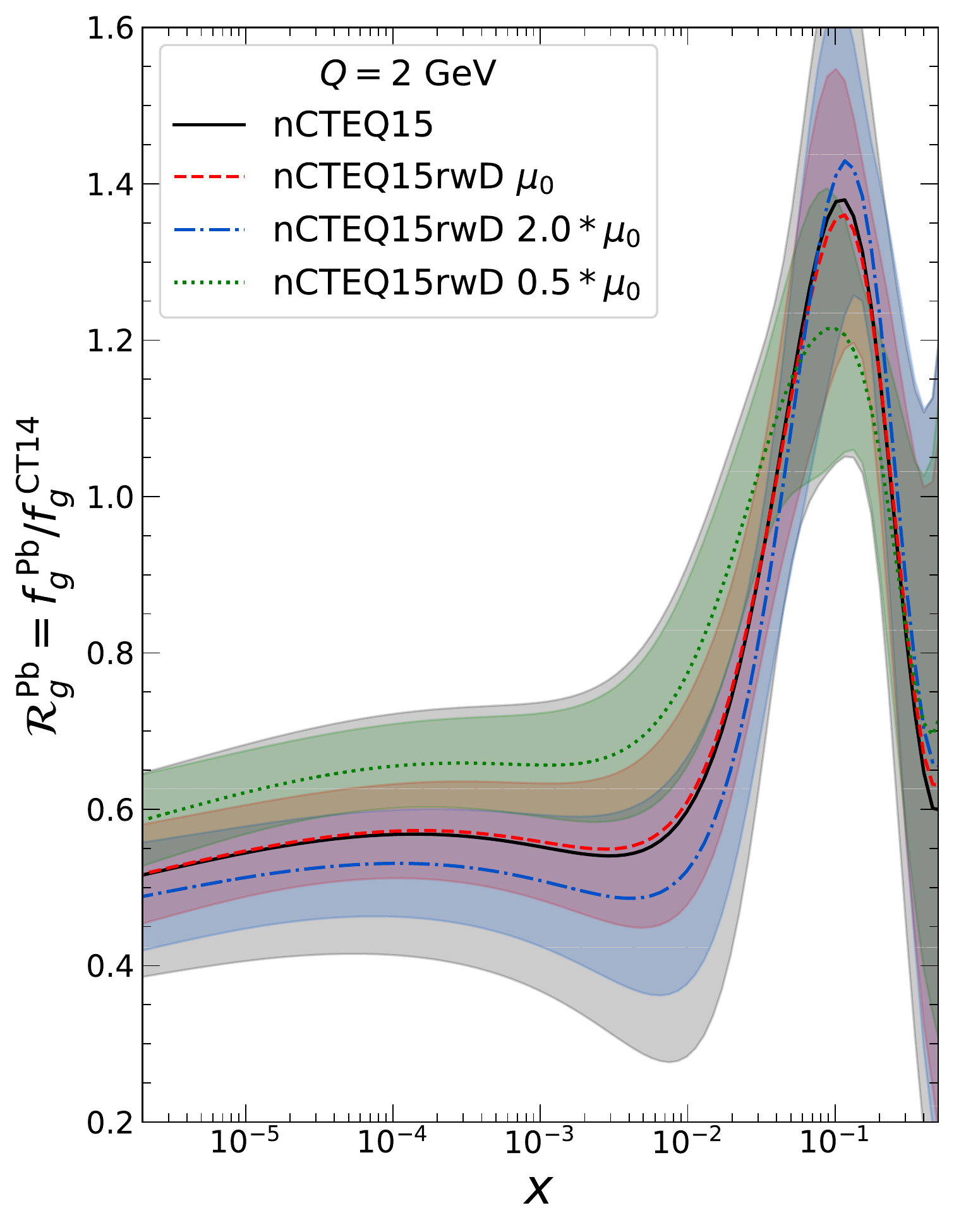}}
\subfloat[$D$-reweighted EPPS16\label{fig:EPPS16compDscale}]{
\includegraphics[width=0.5\columnwidth]{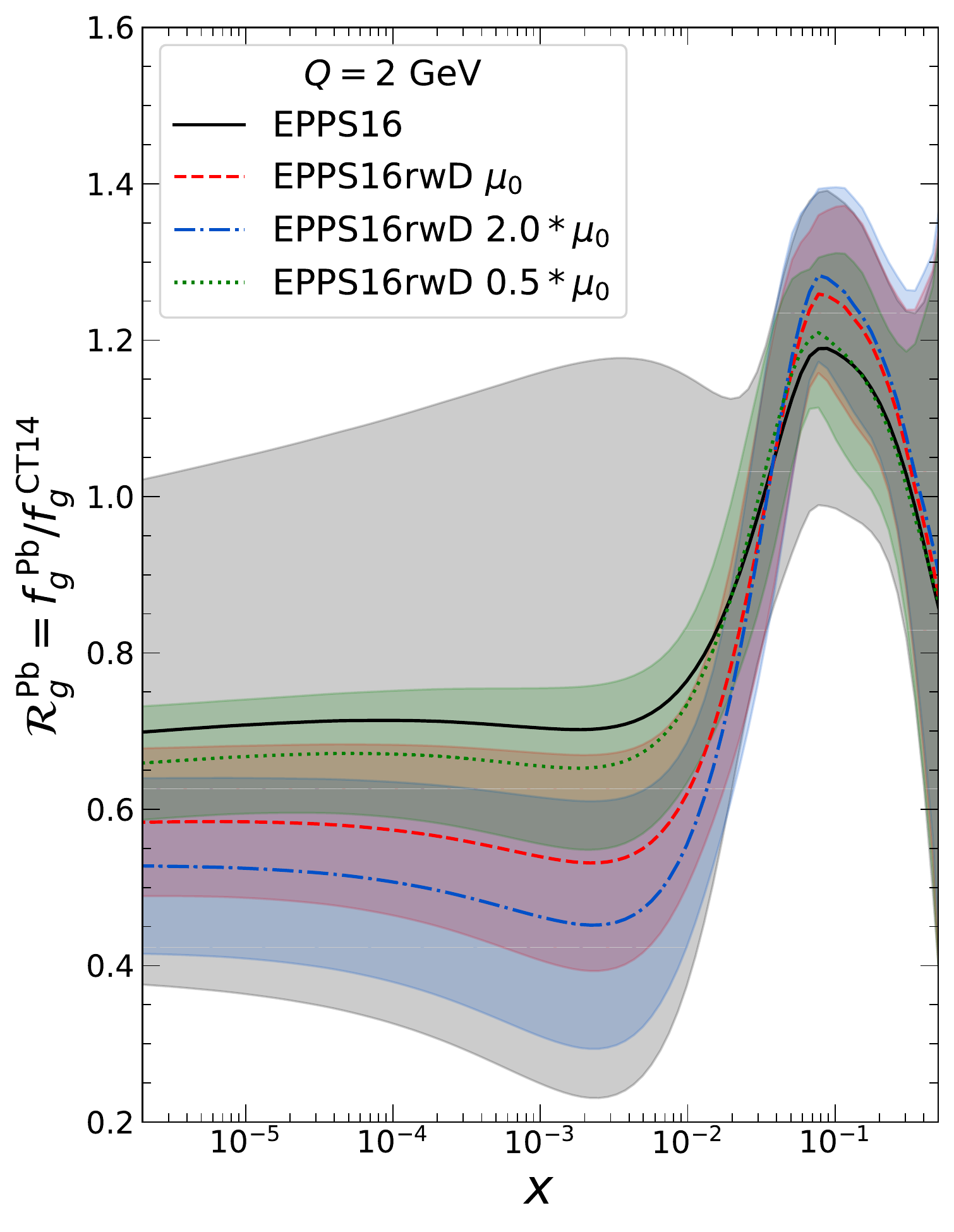}}\\
\subfloat[$B\to J/\psi$-reweighted nCTEQ15\label{fig:nCTEQ15compBscale}]{
\includegraphics[width=0.5\columnwidth]{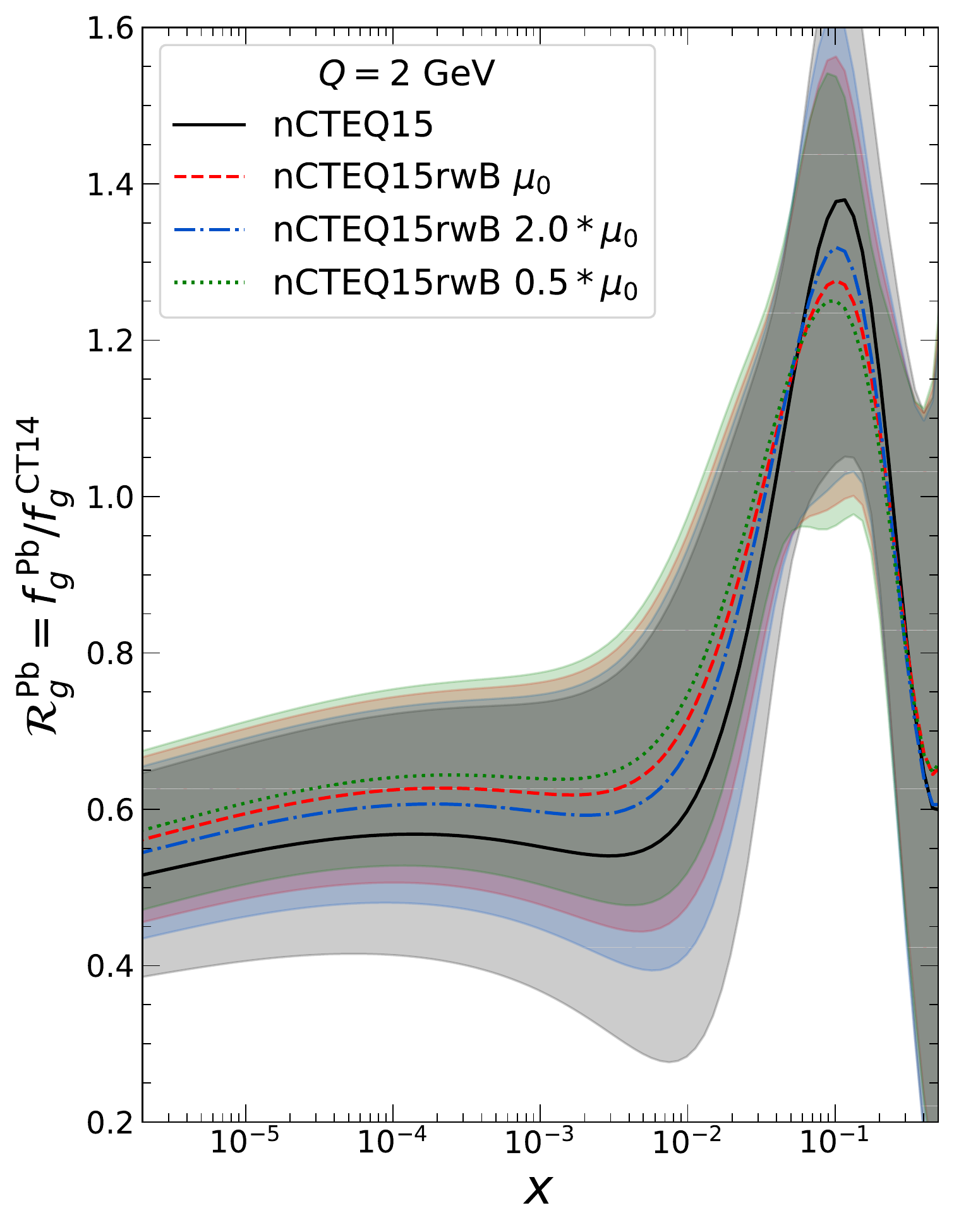}}
\subfloat[$B\to J/\psi$-reweighted EPPS16\label{fig:EPPS16compBscale}]{
\includegraphics[width=0.5\columnwidth]{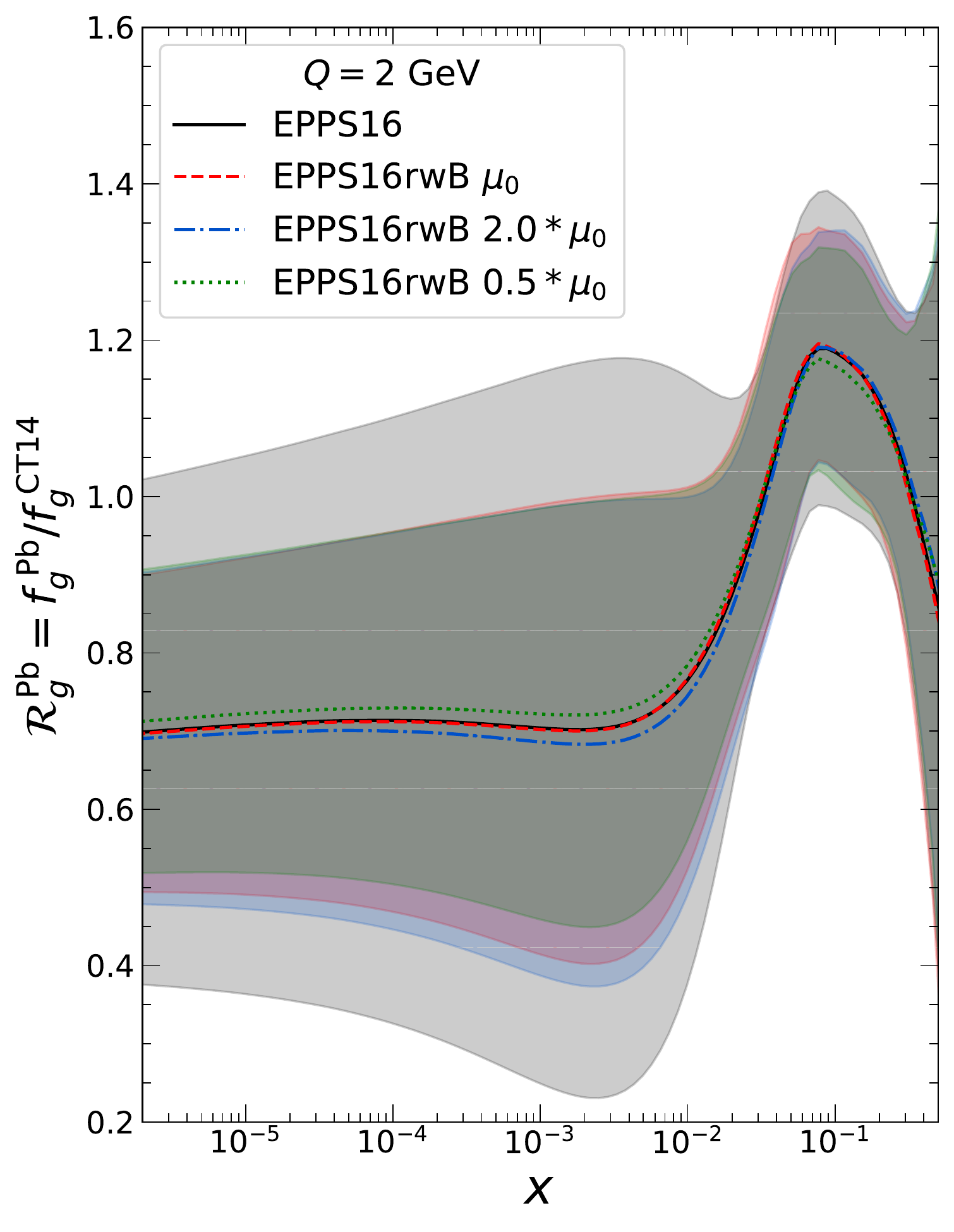}}\\
\subfloat[$J/\psi$-reweighted nCTEQ15\label{fig:nCTEQ15compJpsiscale}]{
\includegraphics[width=0.5\columnwidth]{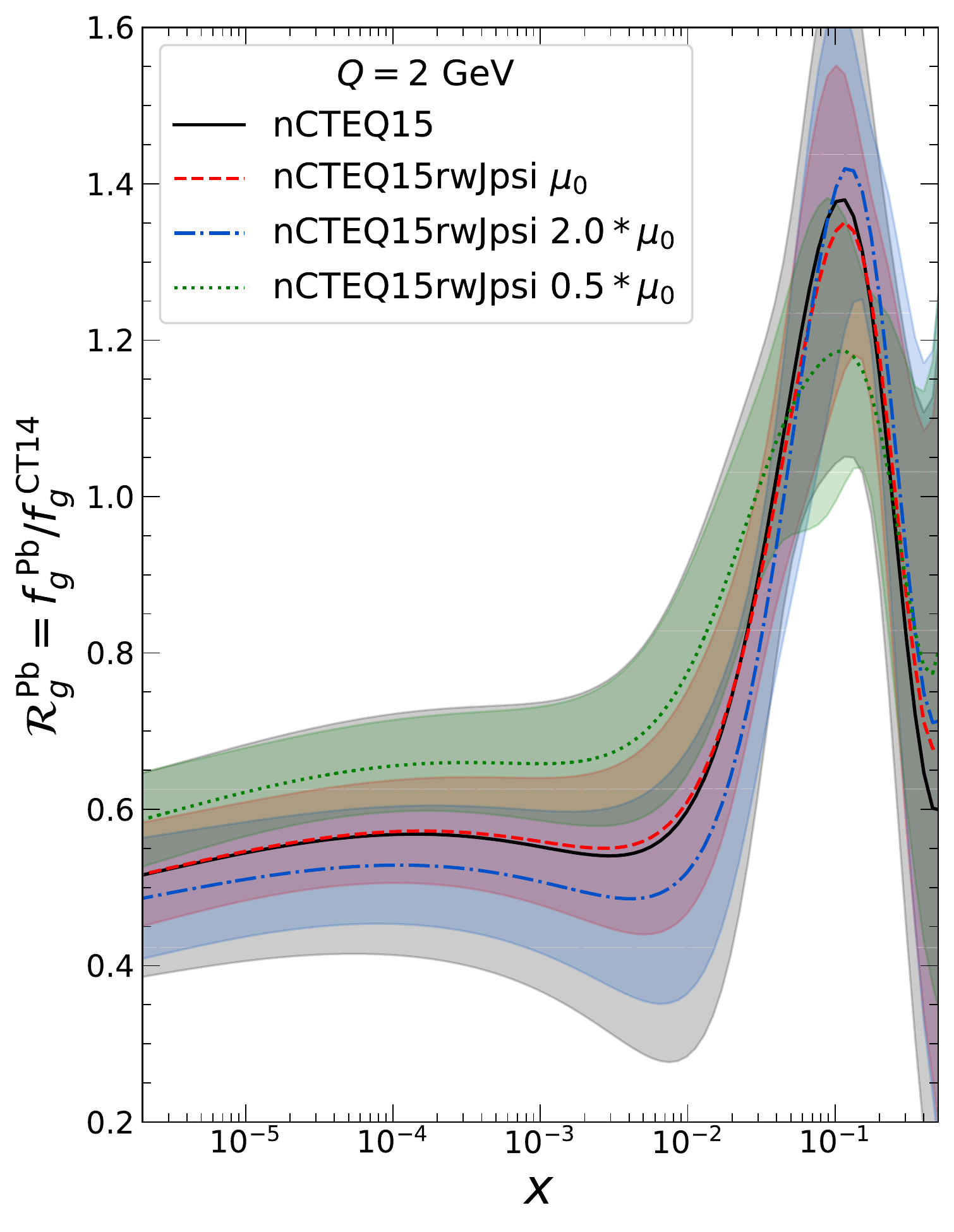}}
\subfloat[$J/\psi$-reweighted EPPS16\label{fig:EPPS16compJpsiscale}]{
\includegraphics[width=0.5\columnwidth]{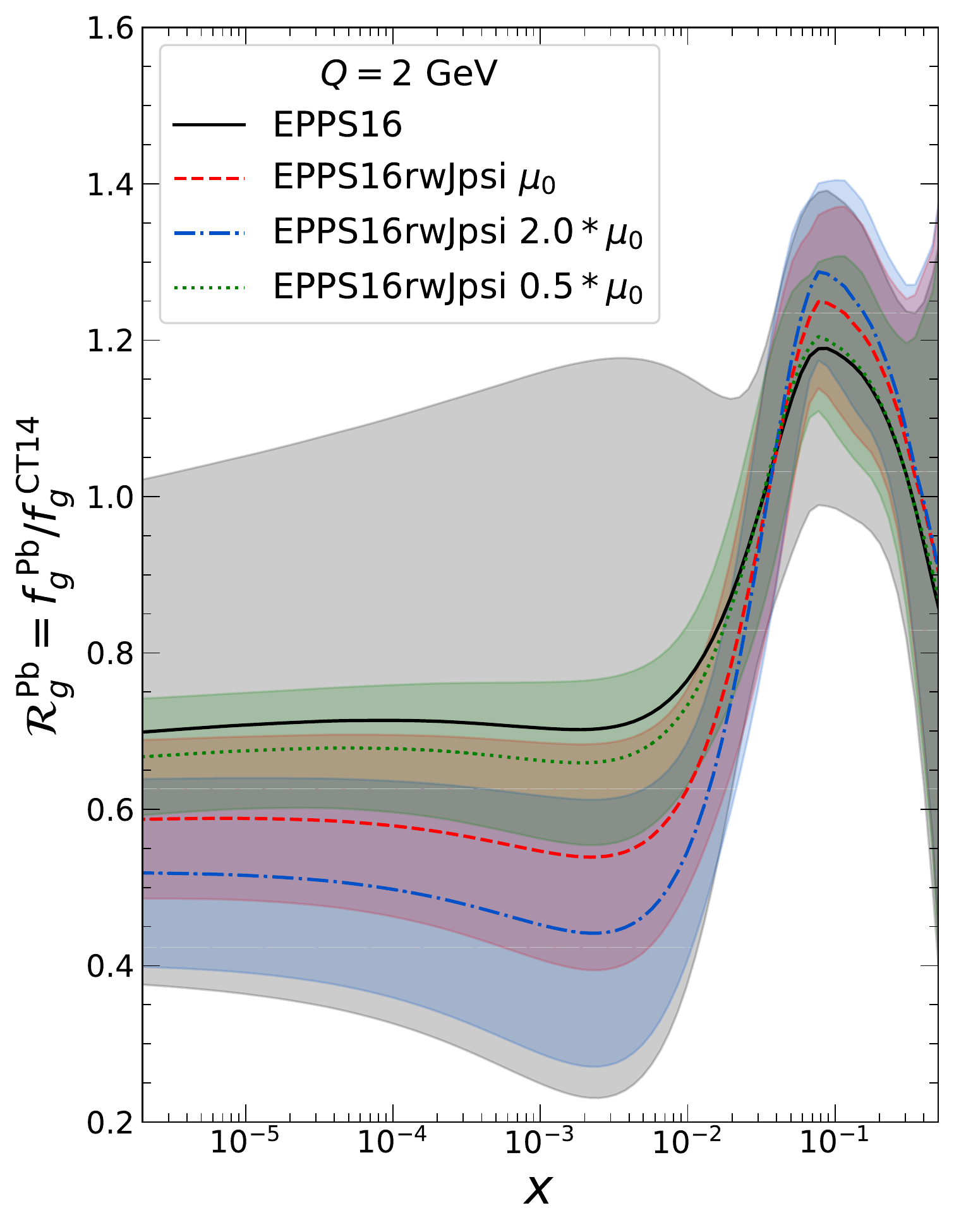}}
\caption{Comparison of the (un)reweighted gluon nuclear ratio
for Pb from the nCTEQ15 (left) and EPPS16 (right) nPDFs 
for reweightings with (from top to bottom):
$D$ meson, $B\to J/\psi$, and $J/\psi$ data. The scale variation was performed
and shown on each plot about $\mu_0$ in Tab.~\ref{tab:scale}).}
\label{fig:PDFcompScale}
\end{figure}

This gives us 3+1 RnPDF sets for each combination of initial nPDF sets
(nCTEQ15 or EPPS16) and each data type which together gives 32 new RnPDF sets.
Since we want to provide all these new RnPDFs to the public such that they
can be used in other studies, we have converted them into corresponding Hessian
sets which are handier to use than the (many) PDF replicas which were initially
obtained.
In order to produce Hessian sets out of PDF replicas we use the {\tt mc2hessian}
program~\cite{Carrazza:2015aoa,mc2hessian}.
For the reweighted Hessian sets we use the same number of error sets as in the
original nPDFs which in case of nCTEQ15 is 32 and in case of EPPS16 is 40.
One should however note that the obtained errors are symmetric and as a result
it is sufficient to provide correspondingly only 16 and 20 error PDFs. The RnPDF
errors for the resulting sets should be computed using the following prescription:
\begin{equation}
\Delta {\cal O} = \sqrt{\sum_k \left( {\cal O}(f_k^{+}) - {\cal O}(f_0) \right)^2 },
\label{eq:errMChess}
\end{equation}
where ${\cal O}$ is a PDF-dependent observable and $k$ goes over 16 or 20 error PDFs.%
    \footnote{The original nCTEQ15 and EPPS16 LHAPDF sets provided both ``plus''
      and ``minus'' error PDFs and as a result the appropriate formula for
      calculating PDF uncertaintiy was
      $\Delta {\cal O} = \frac{1}{2}\sqrt{\sum_k \left( {\cal O}(f_k^{+}) - {\cal O}(f_k^{-}) \right)^2 }$
      instead.}

\begin{figure}[!htb]
\centering{}
\subfloat[$D$  RnPDFs\label{fig:compD}]{
\includegraphics[width=0.7\columnwidth]{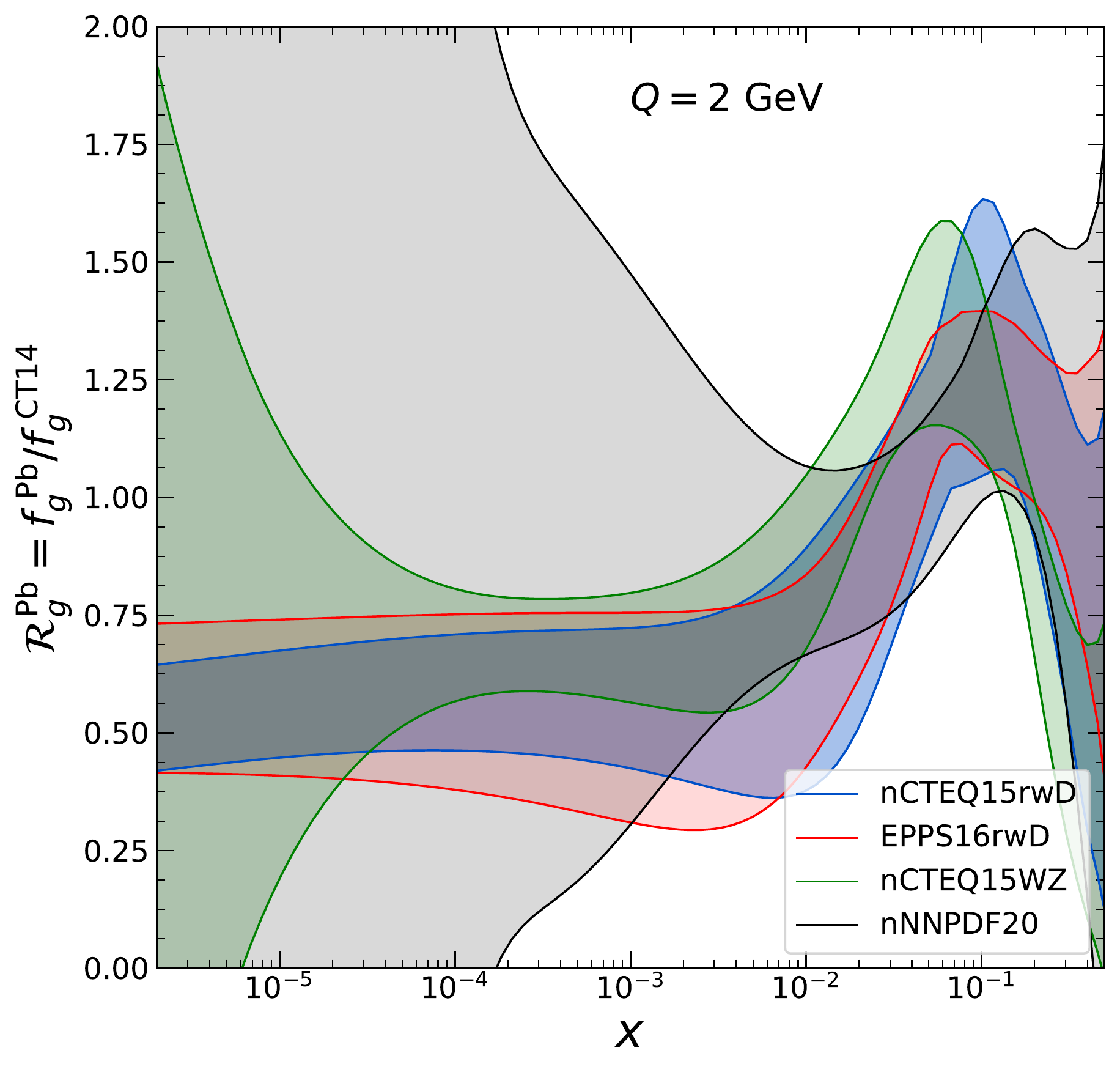}\vspace*{-.2cm}}\\
\subfloat[$B\to J/\psi$ RnPDFs\label{fig:compB}]{
\includegraphics[width=0.7\columnwidth]{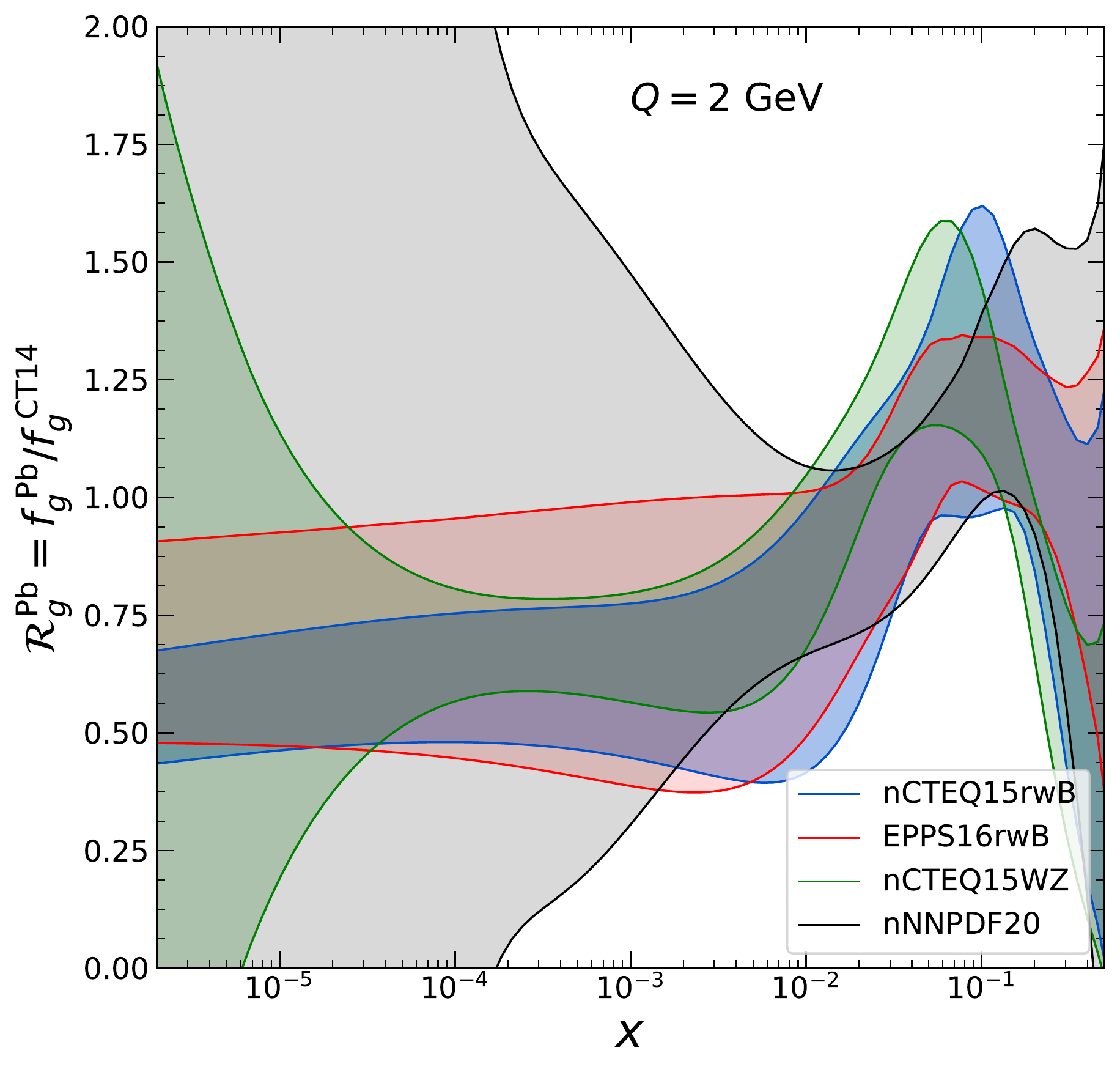}}\vspace*{-.2cm}\\
\subfloat[$J/\psi$ RnPDFs\label{fig:compJpsi}]{
\includegraphics[width=0.7\columnwidth]{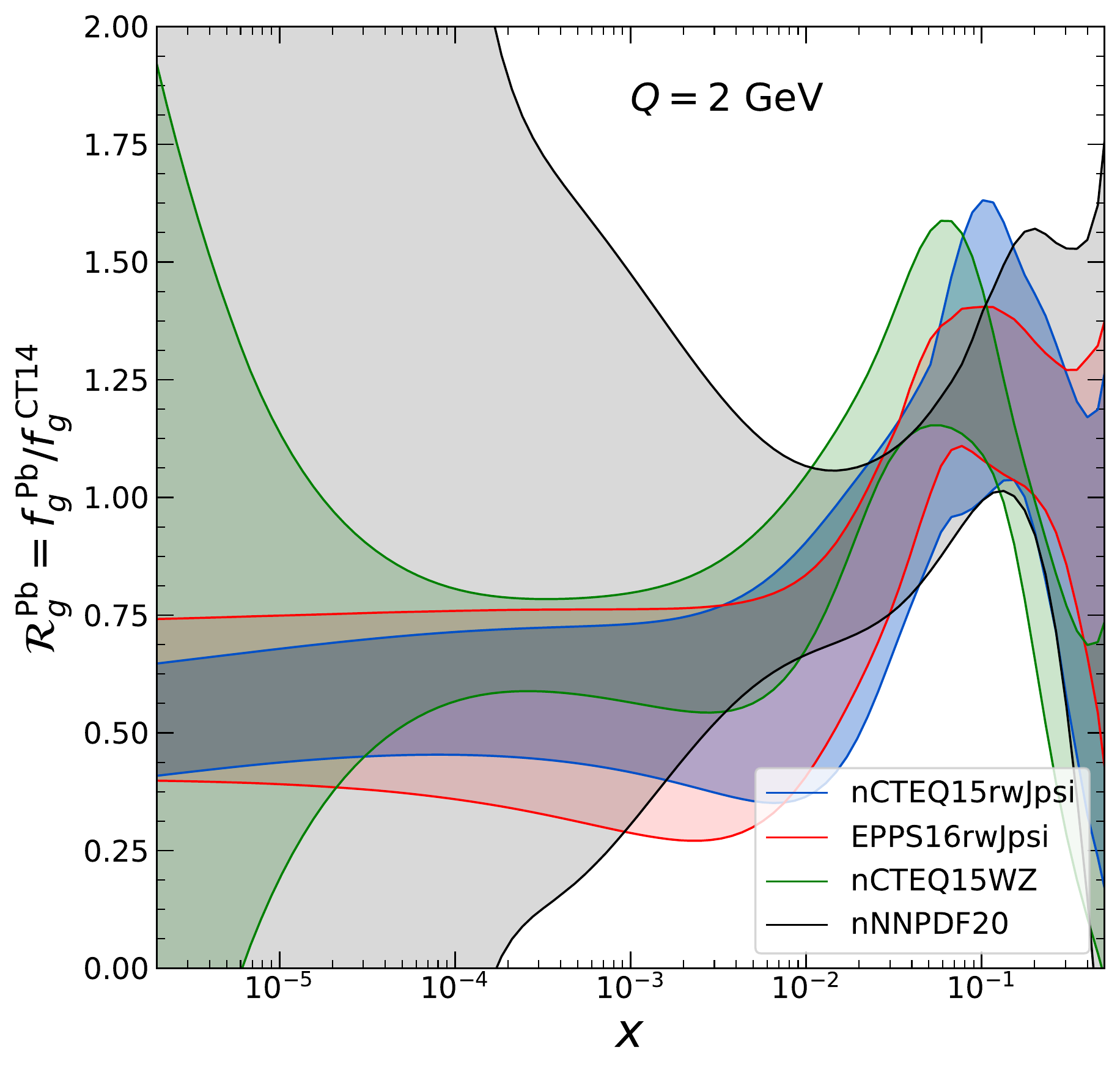}}\vspace*{-.2cm}
\caption{Comparison of the reweighted gluon nuclear ratio for Pb from
the nCTEQ15 and EPPS16 with nCTEQ15WZ~\cite{Kusina:2020lyz}
and nNNPDF20~\cite{AbdulKhalek:2020yuc}.
The ratio is  taken with respect to the CT14nlo proton PDF~\cite{Dulat:2015mca}.
The uncertainties for the RnPDFs are computed as envelopes of uncertainties
for sets obtained with different scales.\vspace*{-.75cm}}
\label{fig:PDFcomp}
\end{figure}

The comparison of the reweighted and original gluon distributions are presented
in Fig.~\ref{fig:PDFcompScale}. We should highlight here that the plots
(and the corresponding LHAPDF files) features PDF uncertainties at 90\% CL.
This is in accordance with the original nPDF sets we used (nCTEQ15 and EPPS16)
but it differs from what we presented in~\cite{Kusina:2017gkz} where
we had used 68\% CL.%
    \footnote{The reweighting itself was performed at 68\% CL but for the
    convenience of the users we have converted the resulting nPDFs to the 90\% CL
    which is the standard used in the community.}
Additional plots showing a detailed comparison of gluon nuclear modifications
for different reweightings both at 68\% CL and 90\% CL are presented in the
appendix~\ref{app:plots}.

In order to confront the resulting distributions with recent nPDFs, in
Fig.~\ref{fig:PDFcomp}, we compare the gluon NMFs
obtained in the HF reweightings with the results from the
nCTEQ15WZ~\cite{Kusina:2020lyz} and nNNPDF20~\cite{AbdulKhalek:2020yuc}
nPDF sets. None of these two nPDFs used the LHC HF data we
employed in our analysis in the fits and, as such, they do not have stringent
constraints on the gluon distribution especially in the low-$x$ region
(since most of the constraints in these sets are coming from the LHC $W/Z$ boson
data which have a kinematic reach to around $x\gtrsim10^{-3}$). Nevertheless,
it is interesting to see the comparison which clear shows that the HF data
is crucial for pinning down the low-$x$ gluon distribution, confirming that
the gluon is shadowed at small values of $x$.

We note that in Fig.~\ref{fig:PDFcomp} we show a single error band for
each of the HF RnPDFs. This error band is an envelope of error
bands originating from the reweightings performed varying the
factorization scale, see Fig.~\ref{fig:PDFcompScale} and
Tab.~\ref{tab:scale}.

Furthermore, since the atomic mass, $A$, of lead and gold nuclei are very close
(208 vs. 197) nuclear modifications for these two nuclei are also very similar.
We used this fact and assumed that the results of reweighting using the $p$Pb
LHC HF data can be directly transferred to the case of gold. To do so,
we simply applied the weights obtained for Pb nPDF replicas to the corresponding Au nPDF replicas. This amounts to assuming that the $A$ dependence in our RnPDFs is the same as that of the original nPDFs. That way, one can see the impact of HF \pPb\ LHC data on Au
nPDFs and confront such new information with $p/d$Au data from RHIC.

To allow others to use these results we provide LHAPDF grid files for the obtained Hessian
sets. We supply the sets for lead and gold nuclei originating from nCTEQ15 and EPPS16
$J/\psi$, $D$, and $B\to J/\psi$ meson data. We refrain from
providing also the sets obtained from reweightings using $\Upsilon(1S)$ data as the
impact of these data on the original nPDFs was marginal.
The LHAPDF grid files will be available at \url{http://nloaccess.in2p3.fr/HF-LHC-RW-2017}.

%% file: predictions-171220.tex
As we explained above, a data selection is necessary to minimize HT effects~\cite{Segarra:2020gtj}. 
In addition, we recall that our parametrization of the hard scattering neglects the (anti-)quark contributions. These are irrelevant at the LHC. However, it is important that one remains in a kinematical region where gluon fusion dominates to apply this parametrization. For charmonia, this remains the case down to  the energy range of RHIC~\cite{Feng:2015cba,Lansberg:2010vq, Brodsky:2009cf} and the LHC in the fixed-target mode~\cite{Hadjidakis:2018ifr,Massacrier:2015qba,Lansberg:2012kf,Brodsky:2012vg} but it is also important to keep away from those regions where other nuclear effects than those which 
can be encapsulated in the nPDFs are believed to be important. 
This explains why we originally restricted our reweighting study to HF production in $pA$ collisions at LHC energies.

We could also have added the forward $d$Au $J/\psi$ RHIC data. Instead, we preferred to focus on the LHC data and to use the RHIC~\cite{Adare:2010fn,Adare:2012qf} and recent LHC~\cite{Aaboud:2017cif,Adam:2016ich} ones as cross checks. We also note that adding the RHIC data would in fact have constrained Au nPDFs. Here, what we rather do is to assume that the relative $A$ dependence of Au and Pb shadowing is the same as in the original nPDFs.

We extend this validation by also showing comparisons with a number of data sets which appeared after our study~\cite{Acharya:2019zjt,Acharya:2018kxc} and which can be considered as predictions since the nPDFs were left unchanged. However, our objective is not to be exhaustive but rather to illustrate that our RnPDFs provide a first good estimate of nuclear effects at work on the production of the corresponding particles and that our released LHAPDF grids can be used as such. Our examples have also been chosen to explain which LHAPDF grids to use and how, that is with which $\mu_F$ scale choices.

We should indeed distinguish two cases. The first corresponds to the situation where a LHAPDF grid has been reweighted on the data of same process  (or a very similar one) as that for which one wishes to provide NMF predictions. In such a case, the $\mu_F$ uncertainties are likely highly correlated even if the $x$ range is not similar.\footnote{We however note that, for low-$P_T$ quarkonia at NLO, the NLO QCD corrections generate a subtle $\mu_F$ vs $x$ interplay which results in negative cross sections at increasing energies for $\mu_F$ larger than the quarkonium mass. In such a case, we recently advocated~\cite{Lansberg:2020ejc} the use of a specific $\mu_F$ scale choice. So far, this has only been addressed for $\eta_Q$ production. Once the $J/\psi$ case is addressed, reweighting of nPDFs with this scale choice would be performed.} As such, we advocate that the NMF should be computed for 3 scale choices [$\xi =(0.5,1.0,2.0)$] by using the RnPDF sets corresponding to the same 3 scale values used in the reweighting and then to take the envelope of the resulting NMF uncertainties. This is what we will show next for $J/\psi$ and charm production.

On the contrary, if the process for which one wishes to compute the NMF is a different one, e.g. di-jet vs charm production, we find it more reasonable to consider that the $\mu_F$ uncertainties are not correlated. In such a case, one should rather use the constraints from the reweighting without any prior and take all the eigensets obtained with the different scale choices on the same footage. Hence, we advocate the use of the new merged grid, the computation of the NMF for 3 scale choices [still $\xi =(0.5,1.0,2.0)$] and the consideation of the envelope of the resulting uncertainties. Since we did not provide LHAPDF grid files for the reweighting with the $\Upsilon$ data, the prediction for the $\Upsilon$ NMFs which we show next are obtained likewise, using the gluon nPDF reweighted with charm data. We have also done so for $B$ predictions. We could have used those from beauty or $J/\psi$ in both cases as well. 

\begin{figure}[hbt!]
\centering
\includegraphics[width=0.9\columnwidth,draft=false]{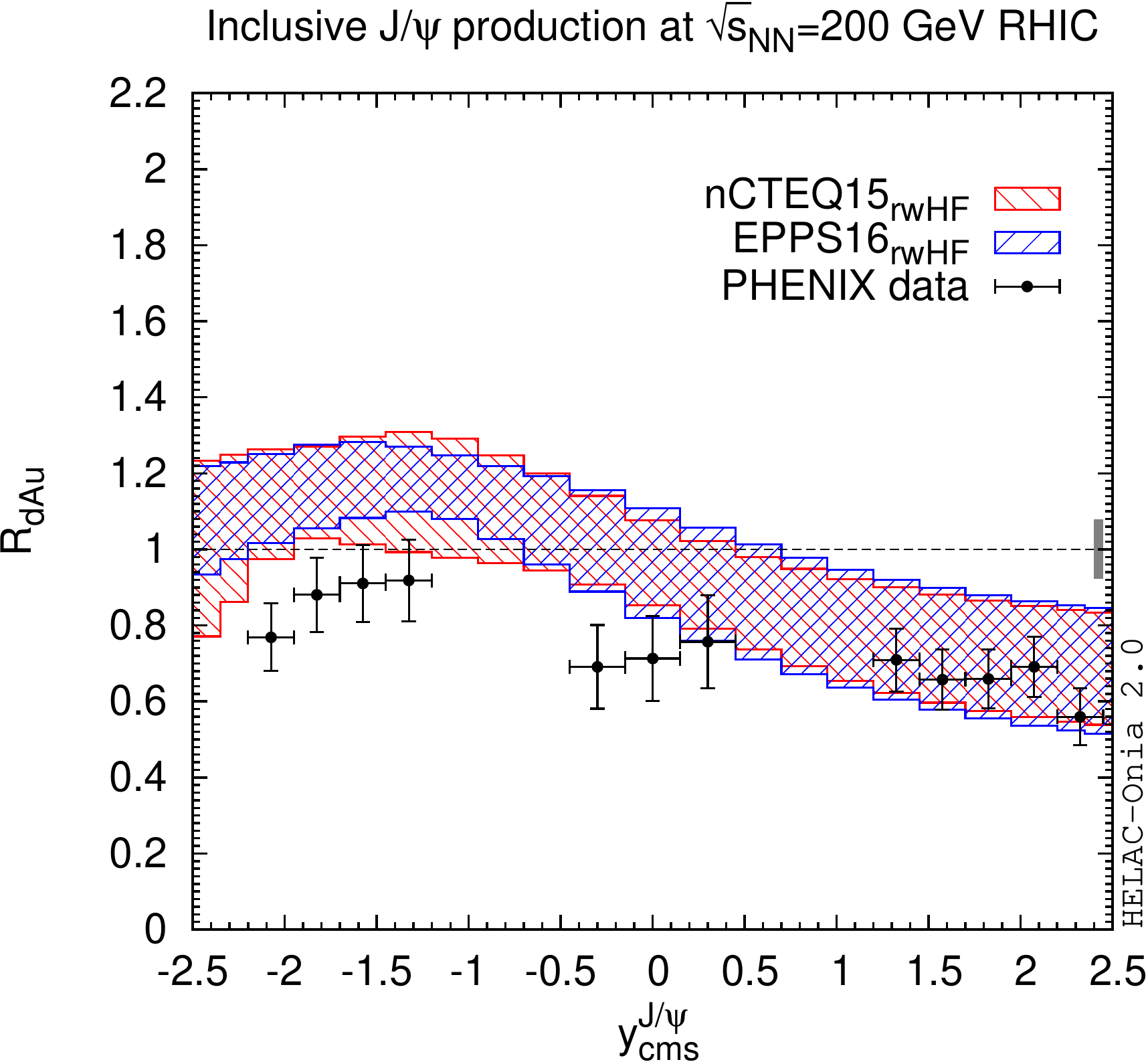} 
\caption{$J/\psi$ \RdAu\ as a function of $\ycms$ at \sqrtsNN=200~GeV computed using our RnPDFs compared to PHENIX data~\cite{Adare:2010fn}. \label{dy_jpsi_RdAu_PHENIXarXiv10101246_rwgthess}\vspace*{-1cm}}
\end{figure}

\begin{figure}[hbt!]
\centering
\includegraphics[width=0.9\columnwidth,draft=false]{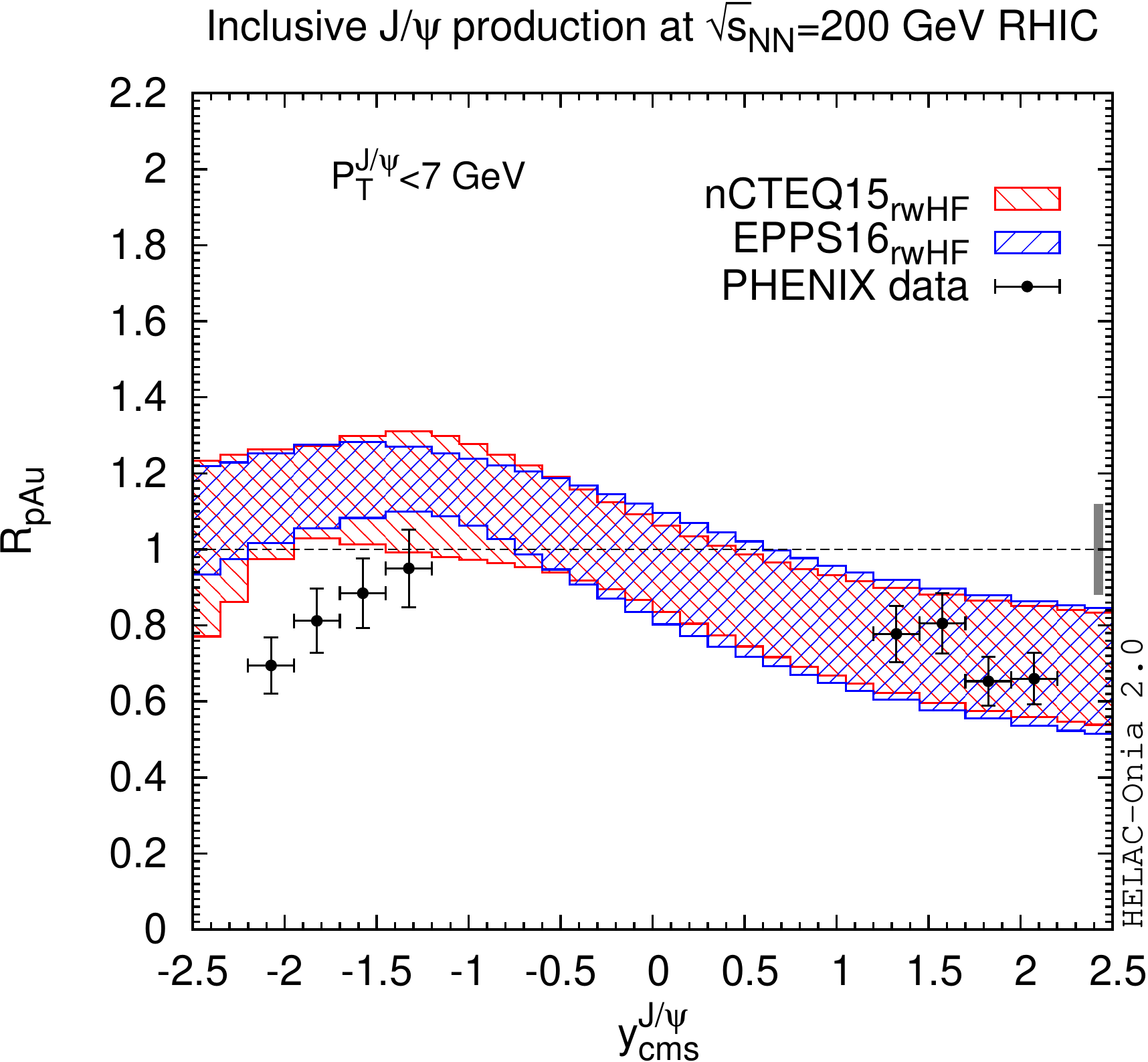}
\caption{$J/\psi$ \RpAu\ as a function of $\ycms$ at \sqrtsNN=200~GeV computed using our RnPDFs compared to PHENIX data~\cite{Acharya:2019zjt}.\label{dy_jpsi_RpAu_PHENIXarXiv191014487_rwgthess}}
\end{figure}

\subsection{$J/\psi$ data at RHIC at \sqrtsNN=200~GeV}

Let us start with some comparisons with RHIC data at $\sqrt{s}=200$~GeV both for $d$Au\footnote{Our results for  $d$Au collisions are obtained by neglecting any nuclear effect in the deuteron.} and $p$Au collisions as a function of $\ycms$ since they illustrate that there is indeed an universality in the suppression of $J/\psi$ forward data at high energies which can be reproduced by nPDF effects only.

\begin{figure}[hbt!]
\centering
\includegraphics[width=0.9\columnwidth,draft=false]{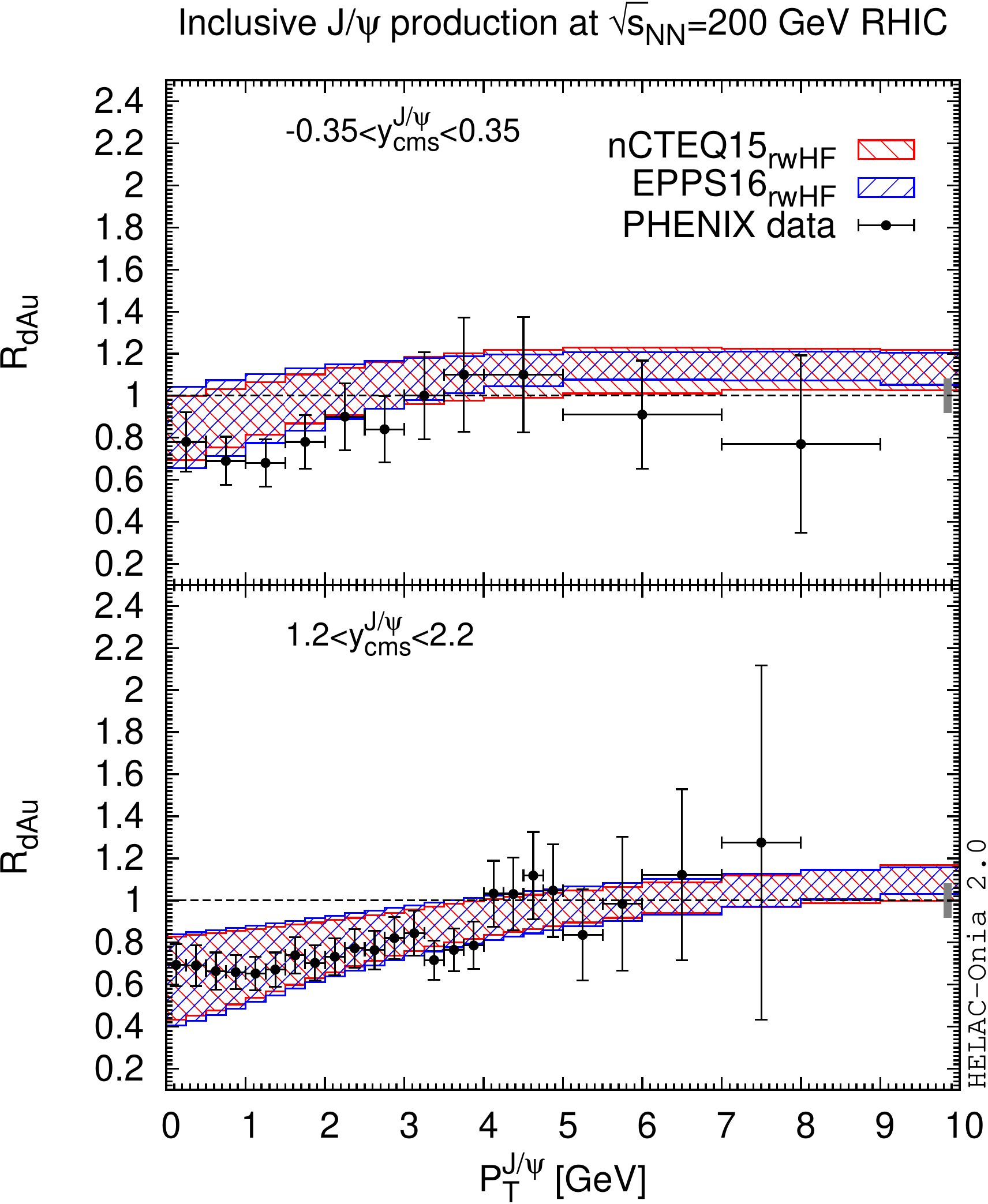} 
\caption{Computed $J/\psi$ \RdAu\ as a function of $\ycms$ at \sqrtsNN=200~GeV using our RnPDFs compared to PHENIX data~\cite{Adare:2012qf}.\label{dpt_jpsi_RdAu_PHENIXarXiv12040777_cen_fwd_rwgthess}}
\end{figure}

\begin{figure}[hbt!]
\centering\includegraphics[width=0.9\columnwidth,draft=false]{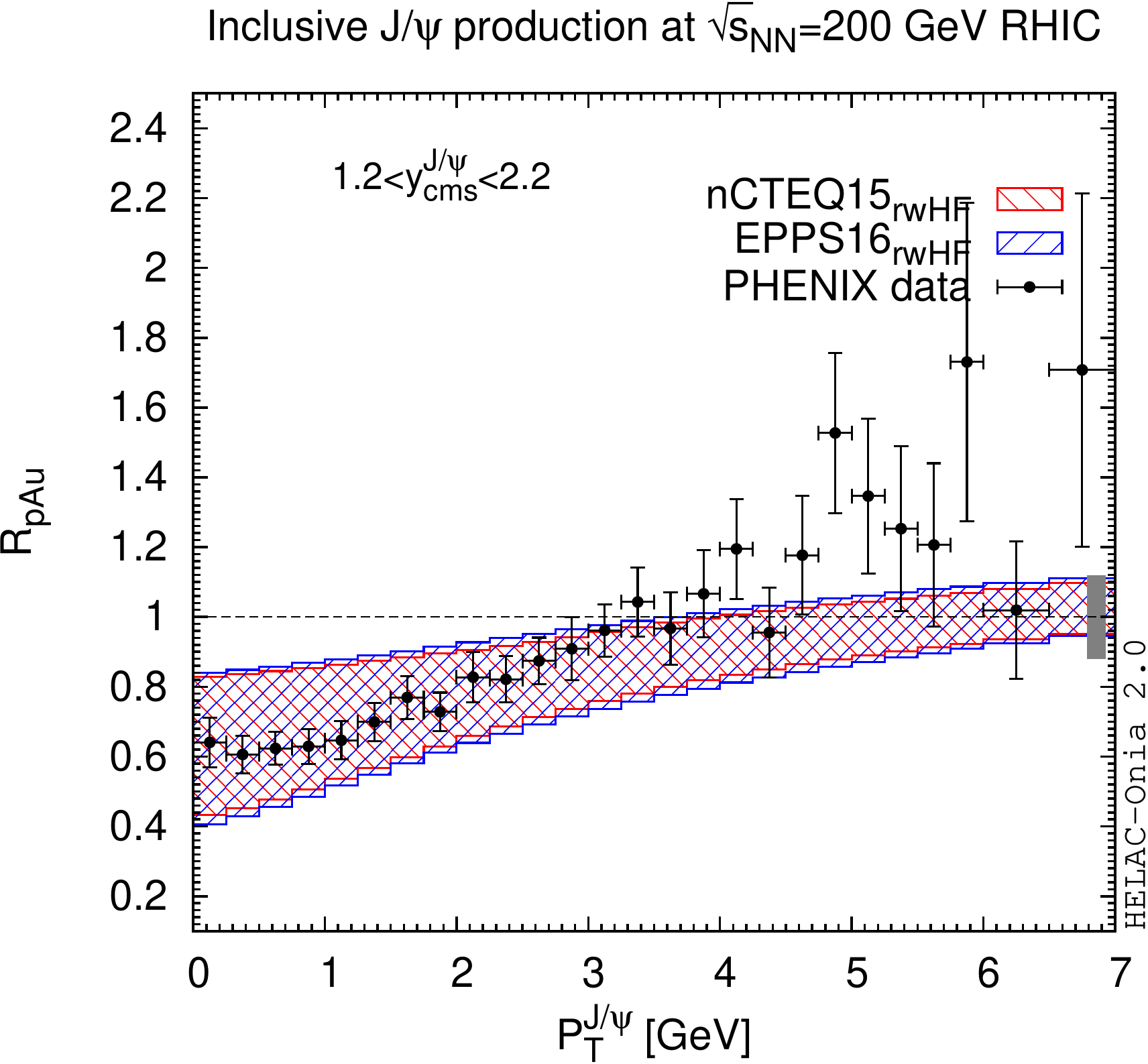}
\caption{Computed $J/\psi$ \RpAu\ as a function of $\ycms$ at \sqrtsNN=200~GeV using our RnPDFs compared to PHENIX data~\cite{Acharya:2019zjt} \label{dpt_jpsi_RpAu_PHENIXarXiv191014487_fwd_rwgthess}\vspace*{-0.5cm}}
\end{figure}

Indeed, one can see in \cf{dy_jpsi_RdAu_PHENIXarXiv10101246_rwgthess} and \cf{dy_jpsi_RpAu_PHENIXarXiv191014487_rwgthess} that the magnitude of the suppression in the forward region is well accounted for by the nPDFs reweighted on the $J/\psi$ LHC data. We recall that the shown uncertainty results from the envelope of the bands found by taking the 3 $J/\psi$-RnPDFs  using 3 scales evaluated at the same 3 scale choices. As we just explained, this amounts to assuming that the factorization scale in the hard scattering is fully correlated between the LHC and RHIC energies. Nevertheless, it remains unknown and should be varied.

On the other hand, one clearly sees that the backward data cannot be described, and to a lesser extent the central-rapidity ones. This is absolutely not surprising --and it is well known that the same happens with the original unreweighted nPDFs~\cite{Lansberg:2016deg}-- since, at these energies and rapidities, the $J/\psi$ has the time to fully form while traversing the nucleus. As such, additional (final-state) effects need to be considered. For instance, they can be encapsulated in an effective absorption. We guide the interested reader to a series of works treating these effects~\cite{Vogt:2004dh,Ferreiro:2008wc,Ferreiro:2009ur,Ferreiro:2012sy}. In what follows, we therefore naturally do not discuss further the $P_T$ dependence of the NMFs in this rapidity region at RHIC. 
\cf{dpt_jpsi_RdAu_PHENIXarXiv12040777_cen_fwd_rwgthess} and \cf{dpt_jpsi_RpAu_PHENIXarXiv191014487_fwd_rwgthess} respectively show
the corresponding \RdAu\ and \RpAu\ vs $P_T$ which are found to be fairly well reproduced by our RnPDFs.

\begin{figure}[h!]
\centering\includegraphics[width=0.9\columnwidth,draft=false]{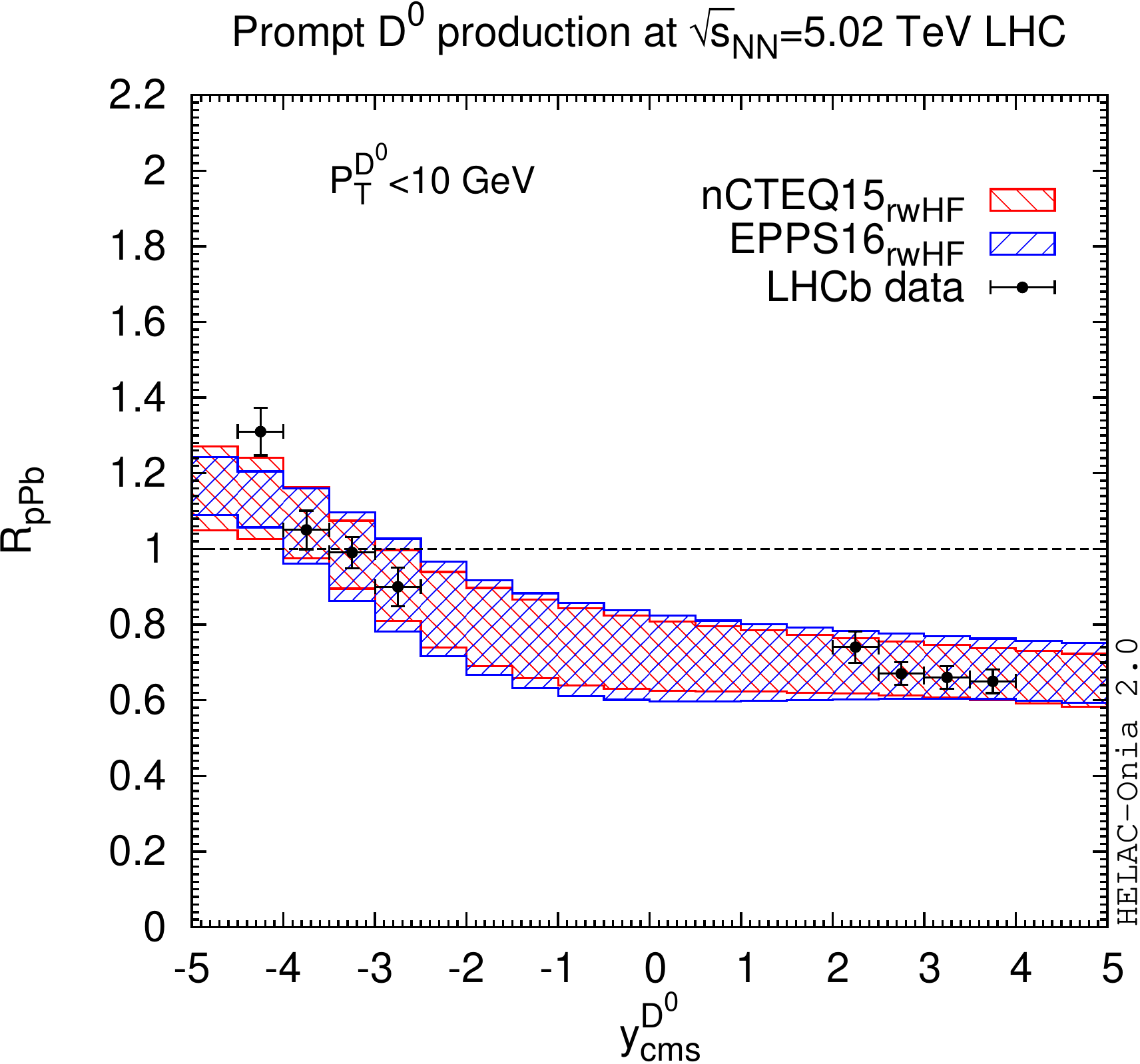}
\caption{Computed $D^0$ \RpPb\ as a function of $\ycms$ at \sqrtsNN=5.02~TeV using our $D$-RnPDFs compared to LHCb data~\cite{Aaij:2017gcy}.\label{dy_D0_RpPb_LHCbarXiv170702750_rwgthess}}
\end{figure}

\begin{figure}[h!]
\centering\includegraphics[width=0.9\columnwidth,draft=false]{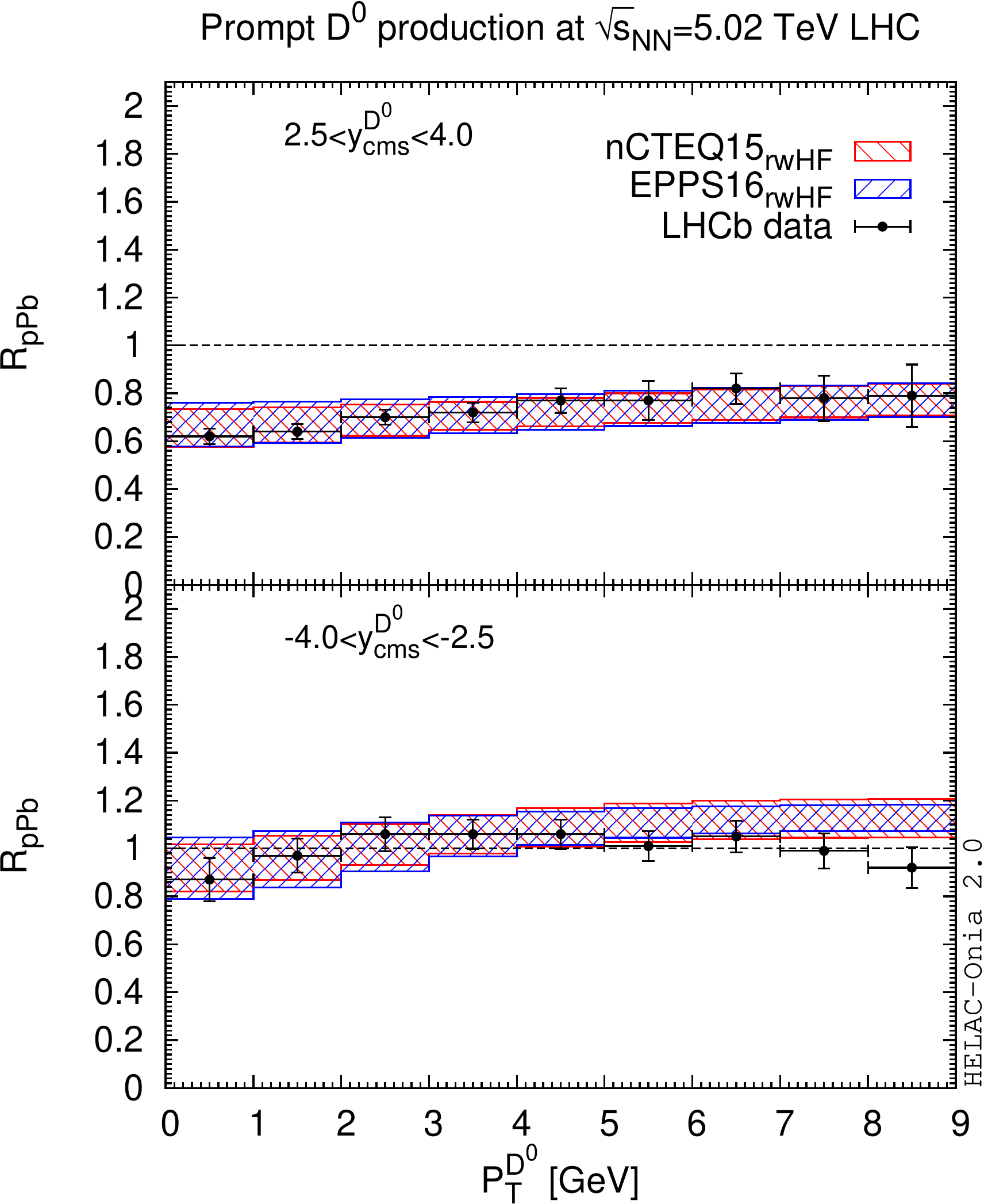}
\caption{$D^0$ \RpPb\ as a function of $P_T$ at \sqrtsNN=5.02~TeV at forward (top) and backward (bottom) $\ycms$ computed using our $D$-RnPDFs compared to LHCb data~\cite{Aaij:2017gcy}.\label{dpt_D0_RpPb_LHCbarXiv170702750_rwgthess}}
\end{figure}

\subsection{$D^0$ at the LHC \sqrtsNN=5.02~TeV}

We start our list of comparisons with LHC data with the $D$-meson ones which are representative of charm data. At this stage, since our examples are mainly illustrative, we have decided not to consider lepton-from-charm data, which are in any case much more complex theory-wise to compute. We recall that our objective  here is not to demonstrate that one can account for the entire set of existing data but to illustrate how and where to use our RnPDF sets and what conclusion to draw from NMFs which would then be obtained.

Our first example is in fact a consistency check for which we show the NMF for $D^0$ at \sqrtsNN=5.02~TeV  obtained from the $D$-RnPDFs precisely using 5.02 TeV data. These have been computed while taking into account the scale correlation as explained above.
\cf{dy_D0_RpPb_LHCbarXiv170702750_rwgthess} shows the same agreement with the \RpPb\ as a function of $\ycms$ measured by LHCb as that obtained in Ref.~\cite{Kusina:2017gkz}. The only difference indeed lies in the procedure to derive the uncertainties with the scale correlation. A similar correspondence is found for the $P_T$ dependence (see \cf{dpt_D0_RpPb_LHCbarXiv170702750_rwgthess}).

\begin{figure}[h!]
\centering\includegraphics[width=0.9\columnwidth,draft=false]{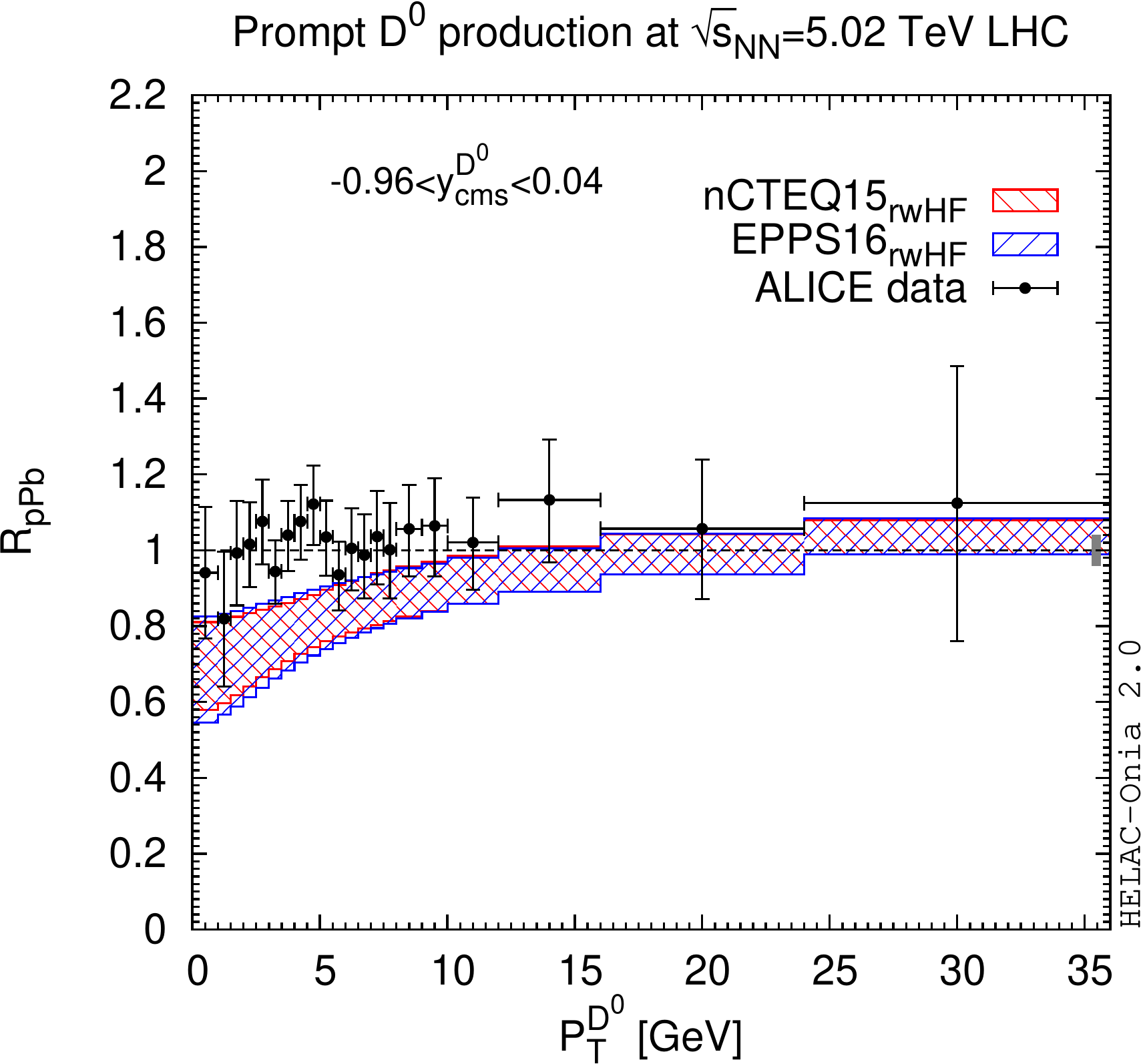}
\caption{$D$ \RpPb\ as a function of $P_T$ at \sqrtsNN=5.02~TeV computed using our $D$-RnPDFs compared to ALICE data~\cite{Acharya:2019mno}.\label{dpt_D0_RpPb_ALICEarXiv19063425_rwgthess}}
\end{figure}

Regarding other $D^0$ data, as of now, we are only able to compare with the central-rapidity data from ALICE as shown on~\cf{dpt_D0_RpPb_ALICEarXiv19063425_rwgthess}. These admittedly exhibit much larger experimental uncertainties and only hint at a possible smaller suppression than what one would expect from our RnPDFs. More precise data at 8.16~TeV  for instance at backward and forward rapidities from LHCb would be welcome, along the lines of their preliminary analysis of $R_{\rm FB}$~\cite{LHCb:2019dpz}.

\begin{figure}[hbt!]
\centering\includegraphics[width=0.9\columnwidth,draft=false]{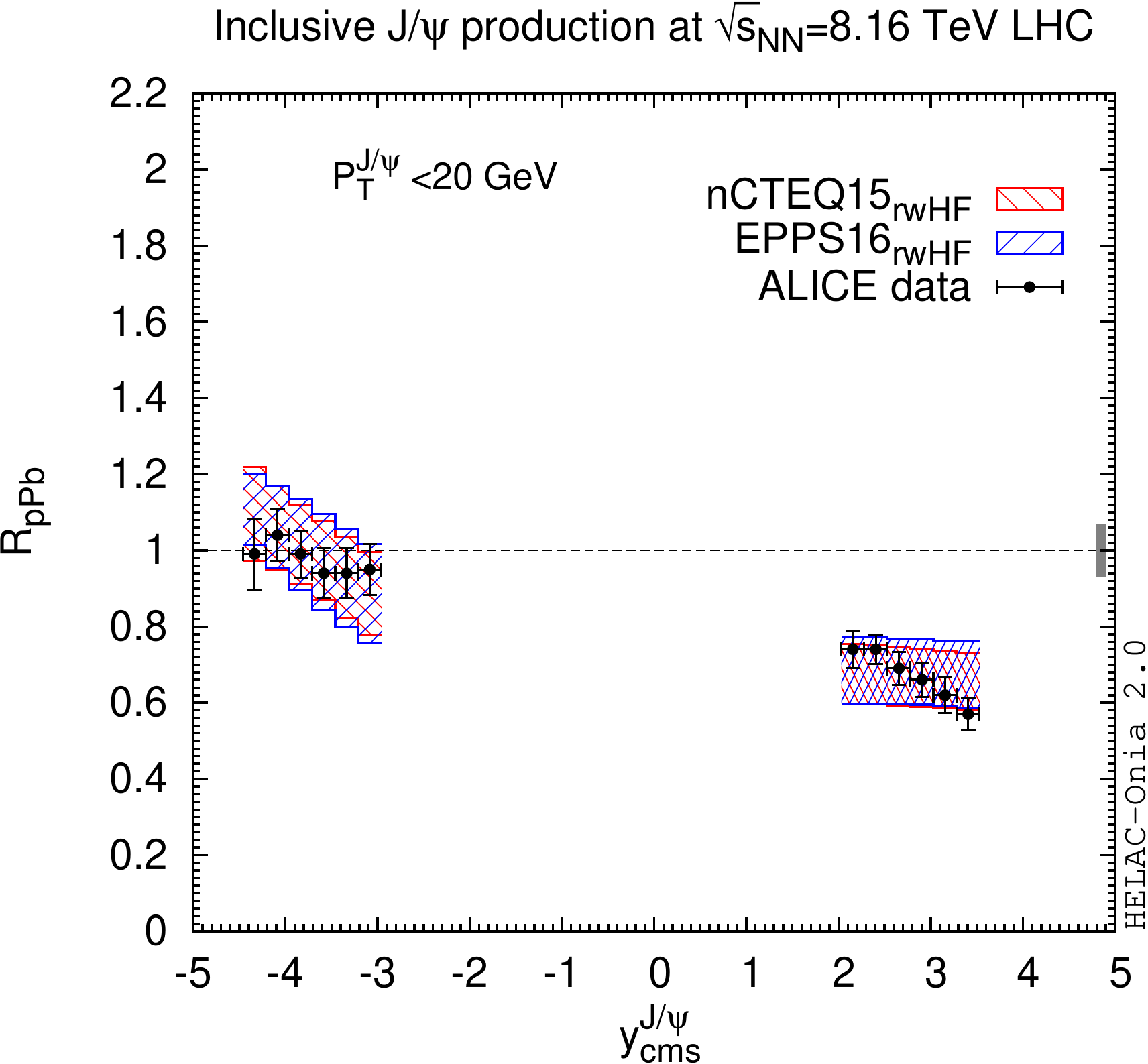}
\caption{$J/\psi$ \RpPb\ as a function of $\ycms$ at \sqrtsNN=8.16~TeV computed using our $J/\psi$-RnPDFs compared to ALICE data~\cite{Acharya:2018kxc}.\label{dy_jpsi_RpPb_ALICEarXiv180504381_rwgthess}}
%
%
\centering\includegraphics[width=0.9\columnwidth,draft=false]{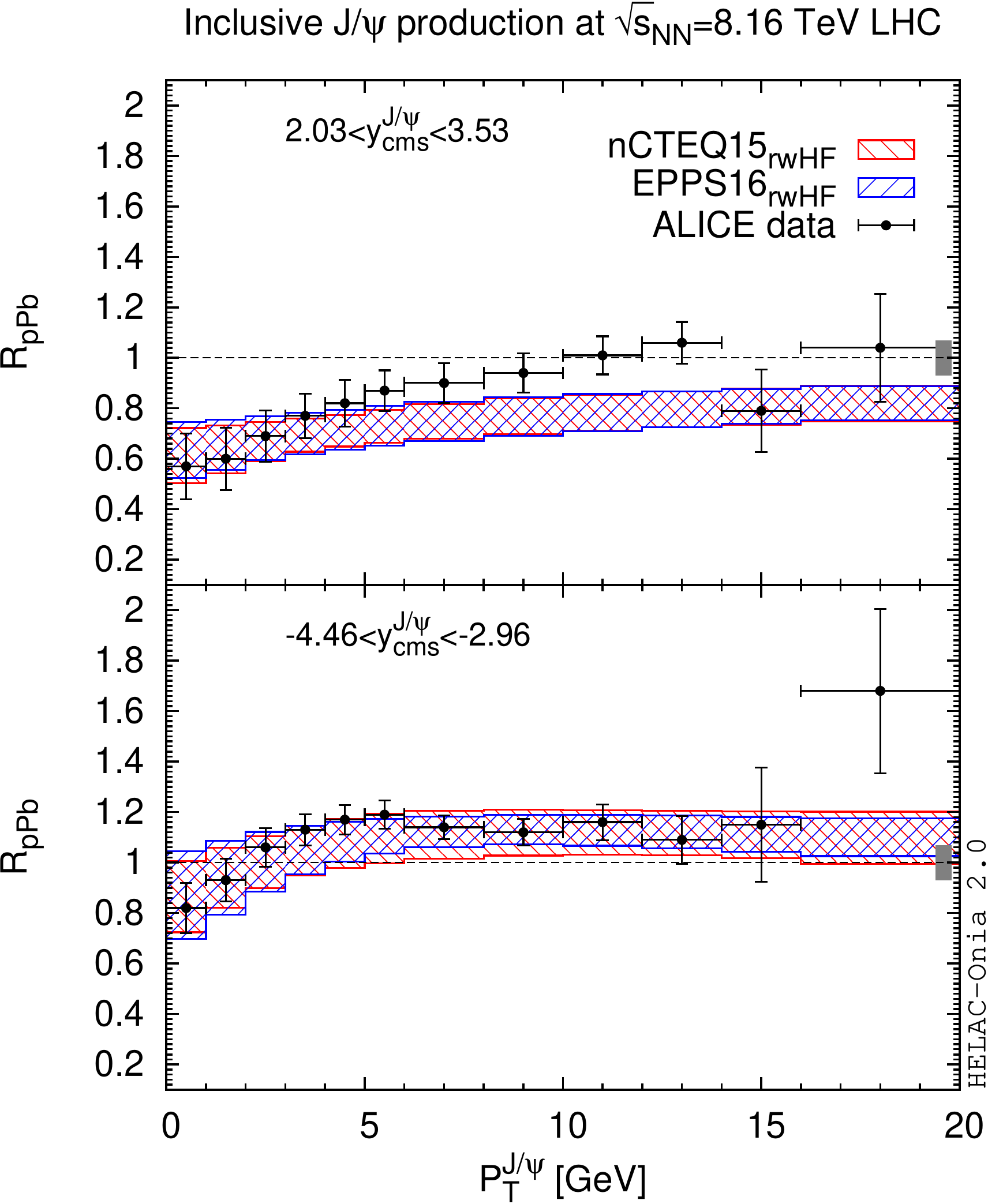}
\caption{$J/\psi$ \RpPb\ as a function of $P_T$ at \sqrtsNN=8.16~TeV  at forward (top) and backward (bottom) $\ycms$ computed using our $J/\psi$-RnPDFs compared to ALICE data~\cite{Acharya:2018kxc}.\label{dpt_jpsi_RpPb_ALICEarXiv180504381_rwgthess}}
\end{figure}

\subsection{$J/\psi$ at \sqrtsNN=8.16~TeV}

We now move on to the $J/\psi$ case. Since we have $J/\psi$-RnPDFs, we have used them while taking into account the scale correlation. In this case, we only show comparisons with ALICE data at 8.16 TeV which we did not include in our initial 
reweighting analysis~\cite{Kusina:2017gkz}. \cf{dy_jpsi_RpPb_ALICEarXiv180504381_rwgthess} shows --without much surprise as well--  that the magnitude of the NMF is well reproduced by
our $J/\psi$-reweighted PDFs.

As for the $P_T$ dependence shown on \cf{dpt_jpsi_RpPb_ALICEarXiv180504381_rwgthess}, it is particularly well accounted for in the backward region, less in the forward region of $P_T$. One could be tempted to attribute this to the growing impact of non-prompt $J/\psi$ for increasing $P_T$. 
Indeed, around $P_T=10$ GeV, this non-prompt fraction has already tripled compared
to the $P_T$-integrated value to reach 30\%.
However, anticipating our $B$ results, it is not obvious that this is the case. Another possible explanation is that the agreement in the backward region is coincidental and comes from the onset of the absorption. In such a case, it could be that, in general, the larger $P_T$ region is not well accounted for by our RnPDFs. Once again, we wish to keep this discussion rather descriptive as final physical conclusions would require a full nPDF fit to see if these data are or not reproducible by LT nPDF effects alone.

\subsection{$B$ at \sqrtsNN=8.16~TeV}

We now come to the $B$ meson case. Because the $B$ data were not precise enough when we performed our reweighting analysis~\cite{Kusina:2017gkz}, the results we obtained only showed marginal differences with the original nPDFs. As such, we find it to be a neat example to illustrate how to use our RnPDFs to compute NMFs for processes which are {\em not} connected to those used in the reweighting. In this case, we have used our $D$-RnPDFs with the combined 3-scale results to evaluate the NMFs for 3 scales and have taken the resulting envelope.

\begin{figure}[hbt!]
\centering\includegraphics[width=0.9\columnwidth,draft=false]{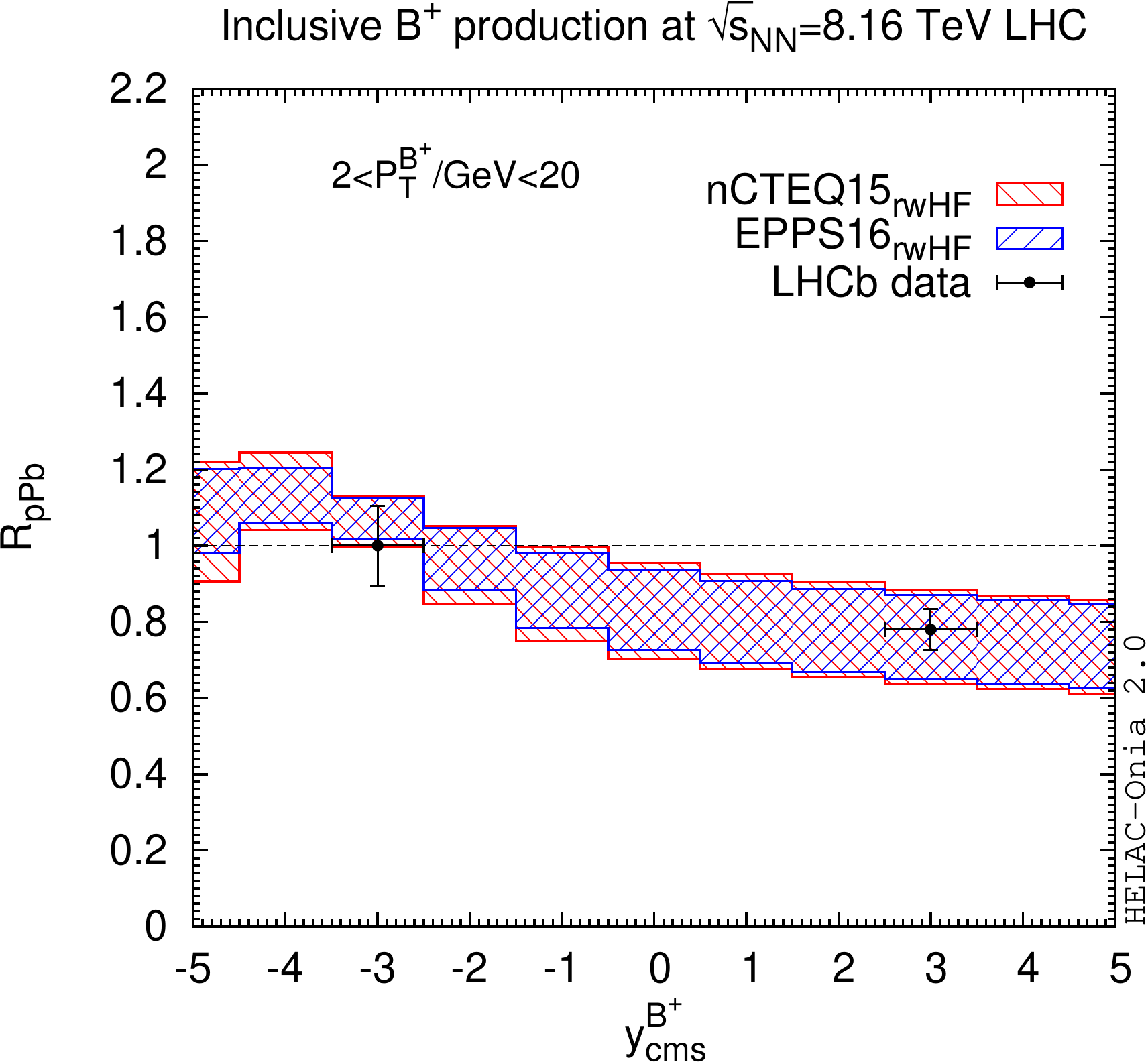}
\caption{$B^+$ \RpPb\ as a function of $\ycms$ at \sqrtsNN=8.16~TeV computed using our $D$-RnPDFs compared to LHCb data~\cite{Aaij:2019lkm}.\label{dy_Bp_RpPb_LHCbarXiv190205599_rwgthess}}
\end{figure}

\begin{figure}[hbt!]
\centering\includegraphics[width=0.9\columnwidth,draft=false]{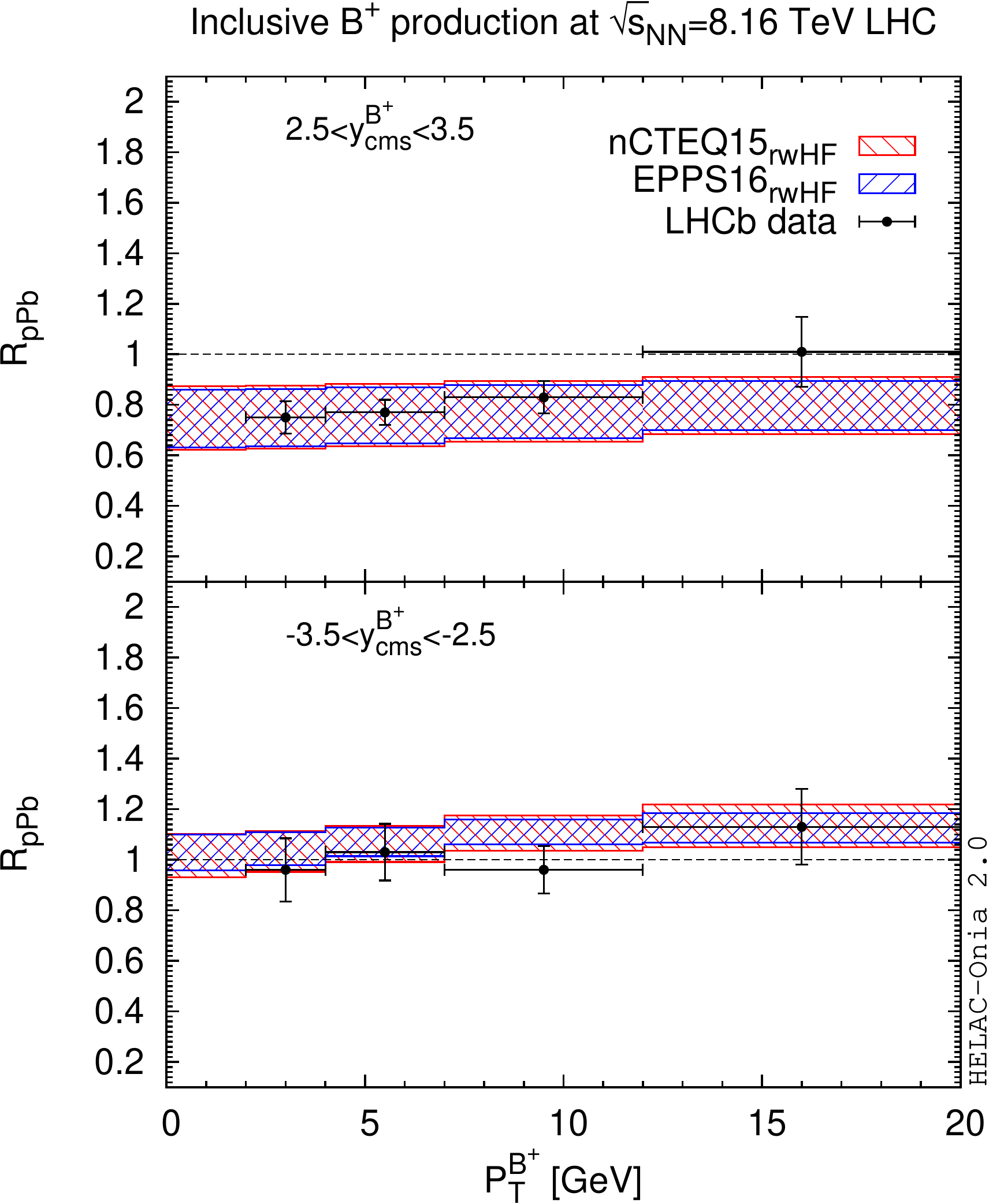}
\caption{$B^+$ \RpPb\ as a function of $P_T$ at \sqrtsNN=8.16~TeV  at forward (top) and backward (bottom) $\ycms$ computed using our $D$-RnPDFs compared to LHCb data~\cite{Aaij:2019lkm}.\label{dpt_Bp_RpPb_LHCbarXiv190205599_rwgthess}\vspace*{-0.5cm}}
\end{figure}

\cf{dy_Bp_RpPb_LHCbarXiv190205599_rwgthess} shows the resulting $\ycms$ dependence which agrees very well with both LHCb $B^+$ data points at 8.16~TeV. We pushed the comparison further with the $P_T$ dependence shown on \cf{dpt_Bp_RpPb_LHCbarXiv190205599_rwgthess} which is in good agreement within the experimental and theoretical uncertainties. We note that we could have performed more such comparisons, with $D$-RnPDF, in particular with $J/\psi$ from $B$ data from LHCb. However, we recall that our objective is certainly not to be exhaustive but to illustrate the usage of our reweigthed nPDFs.

\begin{figure}[hbt!]
\centering\includegraphics[width=0.9\columnwidth,draft=false]{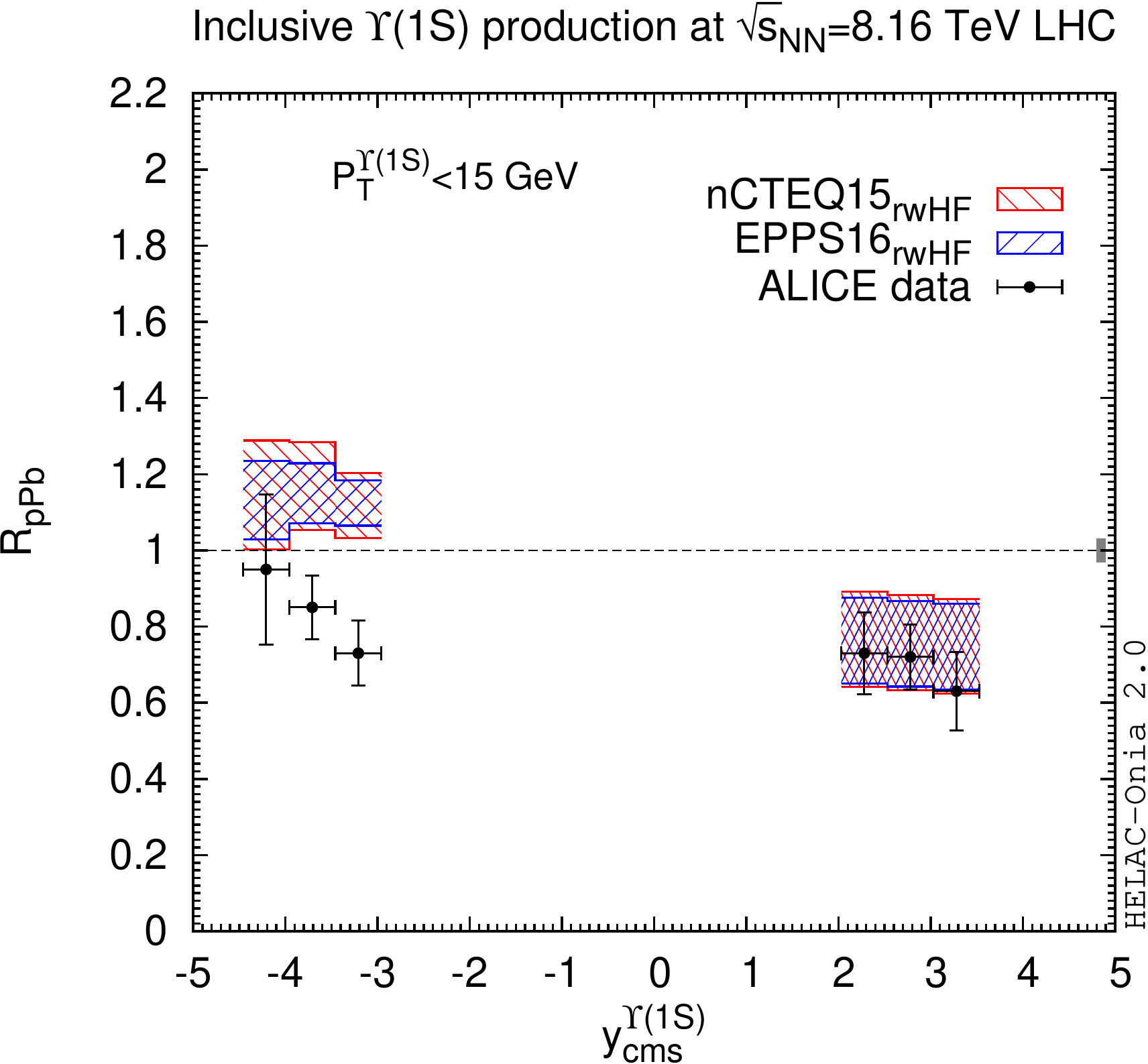}
\caption{$\Upsilon(1S)$ \RpPb\ as a function of $\ycms$ at \sqrtsNN=8.16~TeV computed using our $D$-RnPDFs compared to ALICE data~\cite{Acharya:2020zuq}.\label{dy_Y1S_RpPb_ALICEarXiv191014405_rwgthess}\vspace*{-0cm}}
\centering\includegraphics[width=0.9\columnwidth,draft=false]{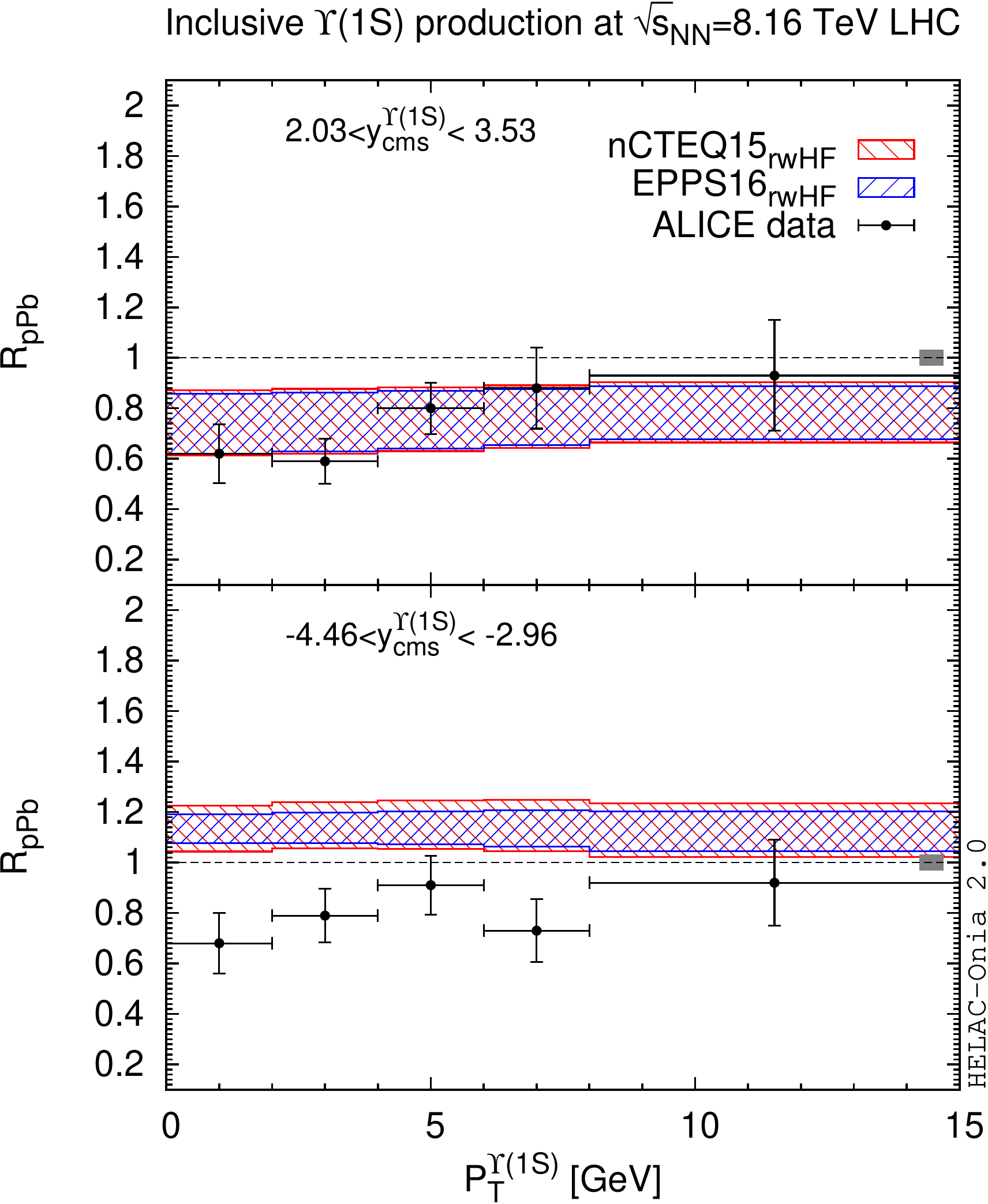}
\caption{$\Upsilon(1S)$ \RpPb\ as a function of $P_T$ at \sqrtsNN=8.16~TeV  at forward (top) and backward (bottom) $\ycms$ computed using our $D$ RnPDFs compared to ALICE data~\cite{Acharya:2020zuq}.\label{dpt_Y1S_RpPb_ALICEarXiv191014405_rwgthess}\vspace*{-0.5cm}}
\end{figure}

\subsection{$\Upsilon(1S)$ at \sqrtsNN=8.16~TeV}

Our last set of comparisons concerns $\Upsilon(1S)$ data. Like for $B$ mesons, we have used $D$-RnPDFs. 
\cf{dy_Y1S_RpPb_ALICEarXiv191014405_rwgthess} shows the resulting $\ycms$ dependence. We find a good agreement in the forward region, but not in the backward region. It seems that the peak generated by the antishadowing is simply absent. Either the $D$ data tend to make it too strong\footnote{We indeed note here that the most backward point for $D$ on \cf{dy_D0_RpPb_LHCbarXiv170702750_rwgthess} is quite high and may have driven the antishadowing in our $D$ RnPDFs too high.}, or the $\Upsilon(1S)$ data are suppressed by another mechanism.\footnote{For instance, comovers could induce a  shift down of \RpPb\ by {0.05}~\cite{Ferreiro:2018wbd}.}
The NMF $P_T$ dependence, shown on \cf{dpt_Y1S_RpPb_ALICEarXiv191014405_rwgthess}, confirms this observation.

\vspace*{-0.5cm}
\subsection{Note on the di-jet data}

After our initial reweighting study came out, the CMS collaboration claimed the first observation~\cite{Sirunyan:2018qel} of a depletion of gluons in Pb at large $x$ based on an analysis of the di-jet yield ratio in $p$Pb and $pp$ collisions as a function of the di-jet pseudo-rapidity, $\eta_{12}$. Such a suppression would correspond to a gluon EMC suppression which was already hinted at by PHENIX $\Upsilon$ data~\cite{Ferreiro:2011xy}. 

It would be very insightful to do a comparison for such ratio using our HF RnPDFs to see if the data constraints concur to the same effect. Yet, we would like to stress that the $pp$ data are not well accounted for by the fixed-order NLO computations and large (positive and negative) $\eta_{12}$. We refer to a neat discussion by Eskola {\it et al.}~\cite{Eskola:2019dui} for more details and a discussion that the deviation may be accounted by modifying the proton PDFs. The discrepancies can easily be on the order of the expected size of the nuclear effects. This happens in the kinematical regions where one would need to look for such a depletion at large $x$ or shadowing at low $x$. 

If we were to bypass a fixed-order analysis by using the same method we have proposed in~\cite{Lansberg:2016deg} by parametrizing the amplitude squared, $|A|^2$, without enough kinematical lever arm in its determination, we would probably hide such a disagreement in $|A|^2$ and our predictions could well be wrong. Yet, in principle, what matters most in our nPDF-based NMF predictions is the relevant $x_1-x_2-Q$ region probed by the scattering. We agree that it is somewhat unlikely that the observed disagreement could be the signal of a phenomenon significantly altering the kinematic of the scattering --this is however not completely excluded if this comes from kinematically enhanced next-to-next-to-leading order (NNLO) QCD corrections as it appears that the di-jet data are better described at NNLO~\cite{AbdulKhalek:2020jut}. Yet, until this issue is settled, we prefer to refrain from performing NMF predictions for dijets. We in fact leave this to the interested reader to do so  since our RnPDFs are now usable by anyone thanks to our released LHAPDF grids. We however suggest a careful reading of~\cite{Eskola:2019dui,AbdulKhalek:2020jut}.

\vspace*{-0.25cm}

%% file: conclusions-171220.tex
We have presented a follow-up of our reweighting study~\cite{Kusina:2017gkz} of two of the recent global fits of 
nuclear parton densities at NLO (nCTEQ15~\cite{Kovarik:2015cma} and EPPS16~\cite{Eskola:2016oht}) which consists
in a release of the corresponding LHAPDF grids with Hessian uncertainties. These will allow anyone to  employ the constraints encoded in the HF experimental data set which we have used for the reweighting in order to perform computations of observables like cross sections or NMFs. 

We had indeed focused on HF LHC data whose predicted yields are however very sensitive to the factorization scale, $\mu_F$. Since the magnitude of the nuclear effects is also strongly $\mu_F$ sensitive, this resulted in a significant dependence of the reweighting on 
the $\mu_F$ we choose. For instance, if one takes a value smaller than the default value, say the transverse mass of the produced particle, the resulting shadowing in the RnPDFs will be weaker. On the contrary, if one reweights with a larger $\mu_F$, 
the resulting RnPDFs will always exhibit a stronger shadowing compared to the former case.  

This naturally induced a significant uncertainty, which we had found~\cite{Kusina:2017gkz} to be as large as that of the resulting $D$ and $J/\psi$-RnPDFs themselves. As we noted, this is less problematic for the bottom-quark sector but the data are not yet precise enough to yield valuable constraints. We have thus generated several RnPDF grids to be used with correlated $\mu_F$ choices if one performs predictions for similar systems as that used for the reweighting. In addition, we release here new RnPDFs with combined scale uncertainty to be used when one assumes no correlation between these scales, \eg\ to predict isolated photon or $\Upsilon$ NMFs from $D$-RnPDFs.

We have thus found it useful to show a selection of comparisons with experimental data to illustrate how to use our RnPDF grids and what to expect from them. In most of the cases, we find a very good agreement with LHC data. It is of course expected in the case of the 8 TeV $J/\psi$ data since one set by LHCb was already included in our reweighting. For the other 8 TeV data, the agreement indicates that the $x$ dependence of the RnPDFs correctly captures the energy dependence of the \RpPb\ and, to some extent as well, highlights the coherence between different LHC data sets. In the case of 200 GeV RHIC data, the agreement we have obtained is even more striking and indeed shows the $x$ dependence of the RnPDFs provide a good description of the gluon distribution up to the upper end of the shadowing region.

Before a full NLO fit using these data is performed, we believe that our RnPDFs can safely be used when the conventional nPDFs show too large uncertainties preventing any physics conclusions. Yet, one should keep in mind that if a strong disagreement is found, it will always be necessary to wonder if a new fitting procedure, which by construction will show more freedom to describe different observables, would not yield a global description including these new data with an acceptable global $\chi^2$. As such our released RnPDFs should be considered as useful and handy tools for observables which are known to be well accounted for by the effects encoded in nPDFs as well as useful exploratory tools for new ones.

%% file: LHAPDFfiles-211220.tex
\section{Cross checks of the reweighted PDFs}
\label{app:plots}
\vspace*{-0.25cm}

\begin{figure*}[!htb]
\centering{}
\subfloat[nCTEQ15\label{fig:ncteq_D_68CL_vs90CL}]{
\includegraphics[width=0.90\textwidth]{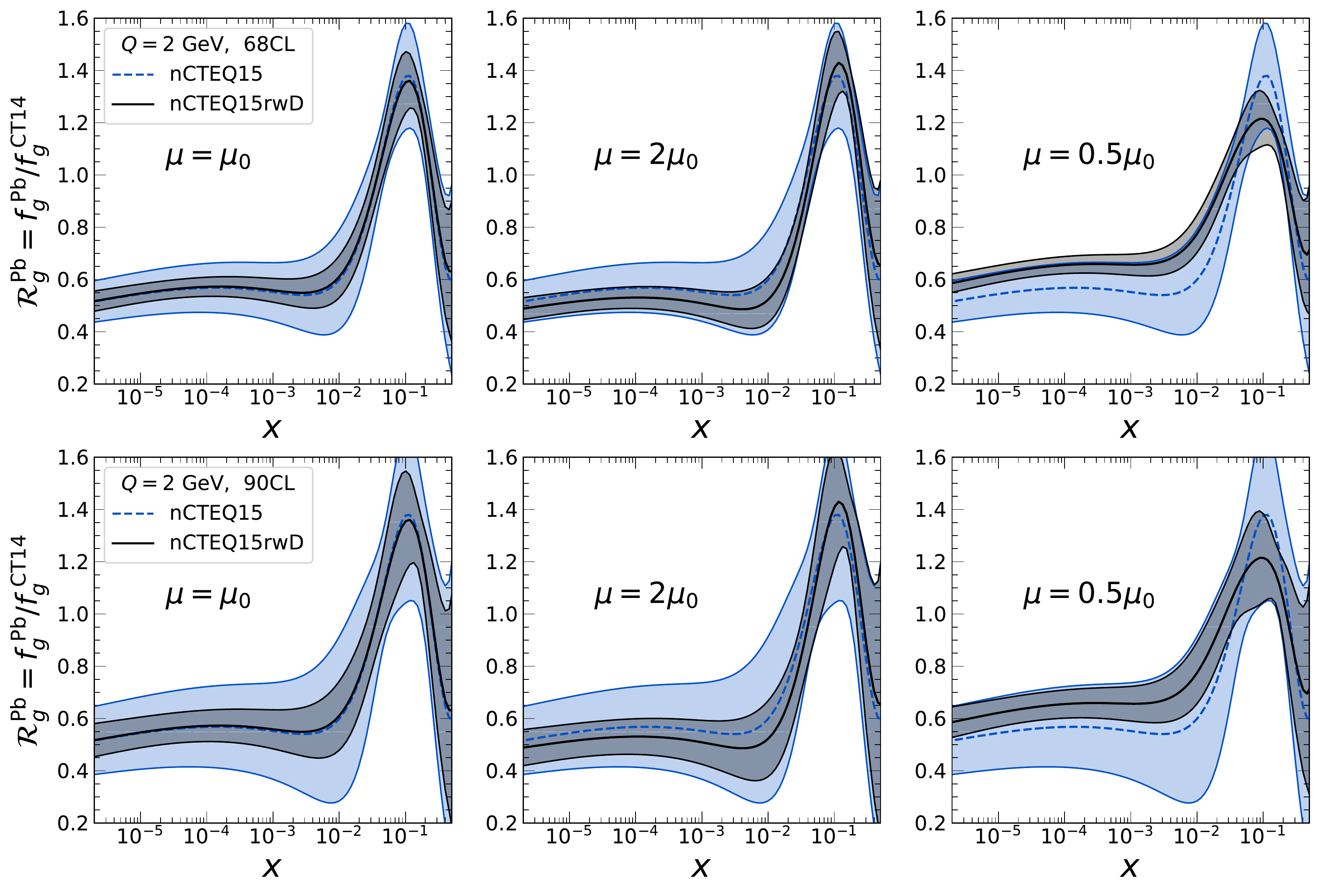}}
\\
\subfloat[EPPS16\label{fig:epps_D_68CL_vs90CL}]{
\includegraphics[width=0.90\textwidth]{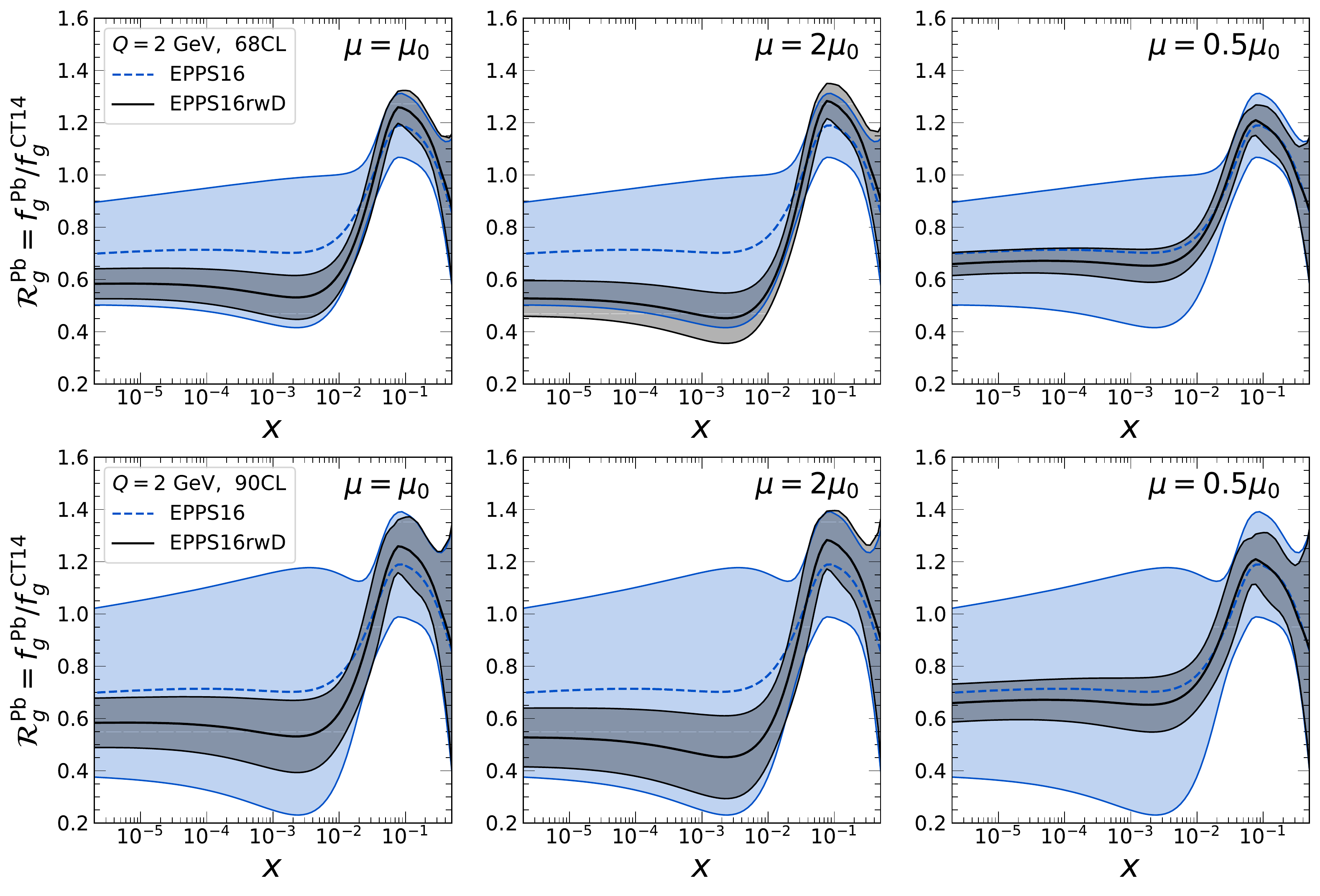}}
\caption{Gluon distribution resulting from reweighting of (a) nCTEQ15
  and (b) EPPS16 nPDFs with $D$-meson data. The upper rows show errors at 68\% CL
  for comparison with Fig.~1f in the original reweighting paper~\cite{Kusina:2017gkz}.
  The lower rows show the same distributions with errors at 90\% CL.}
\label{fig:D_68CL_vs90CL}
\end{figure*}

\begin{figure*}[!htb]
\centering{}
\subfloat[nCTEQ15\label{fig:ncteq_B_68CL_vs90CL}]{
\includegraphics[width=0.90\textwidth]{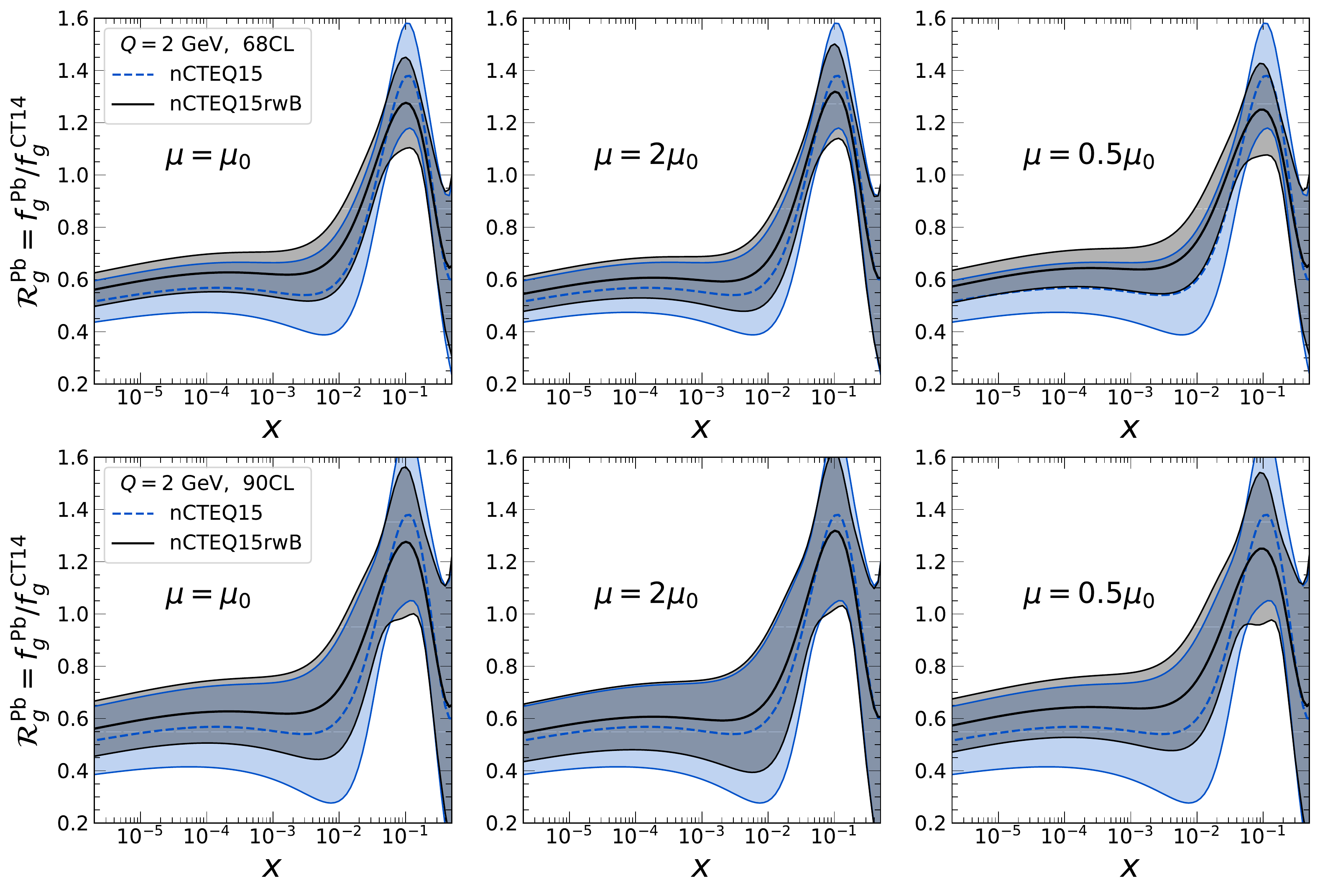}}
\\
\subfloat[EPPS16\label{fig:epps_B_68CL_vs90CL}]{
\includegraphics[width=0.90\textwidth]{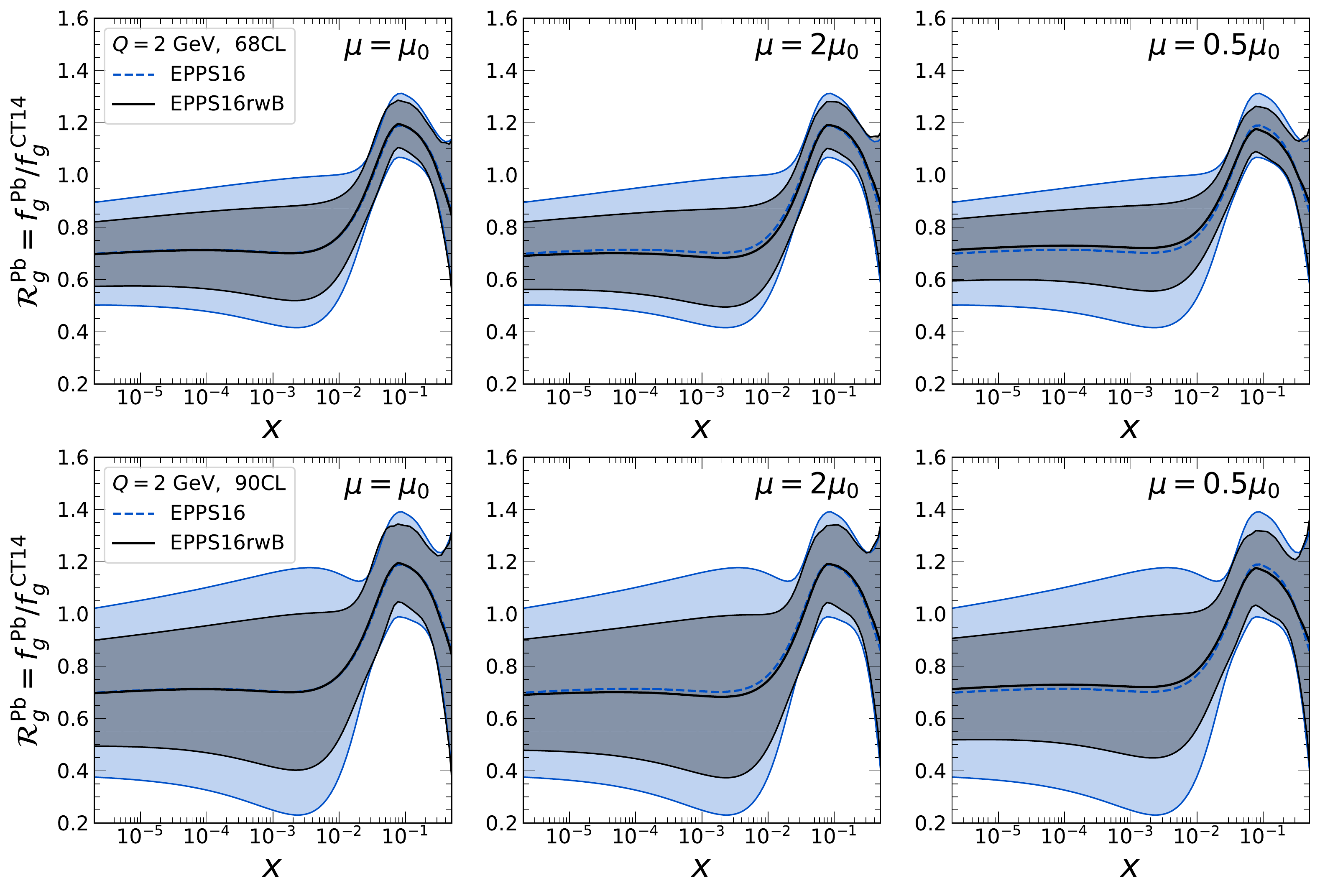}}
\caption{Gluon distribution resulting from reweighting of (a) nCTEQ15
  and (b) EPPS16 nPDFs with $B\to J/\psi$ data. The upper rows show errors at 68\% CL
  for comparison with Fig.~1f in the original reweighting paper~\cite{Kusina:2017gkz}.
  The lower rows show the same distributions with errors at 90\% CL.}
\label{fig:B_68CL_vs90CL}
\end{figure*}

\begin{figure*}[!htb]
\centering{}
\subfloat[nCTEQ15\label{fig:ncteq_Jpsi_68CL_vs90CL}]{
\includegraphics[width=0.90\textwidth]{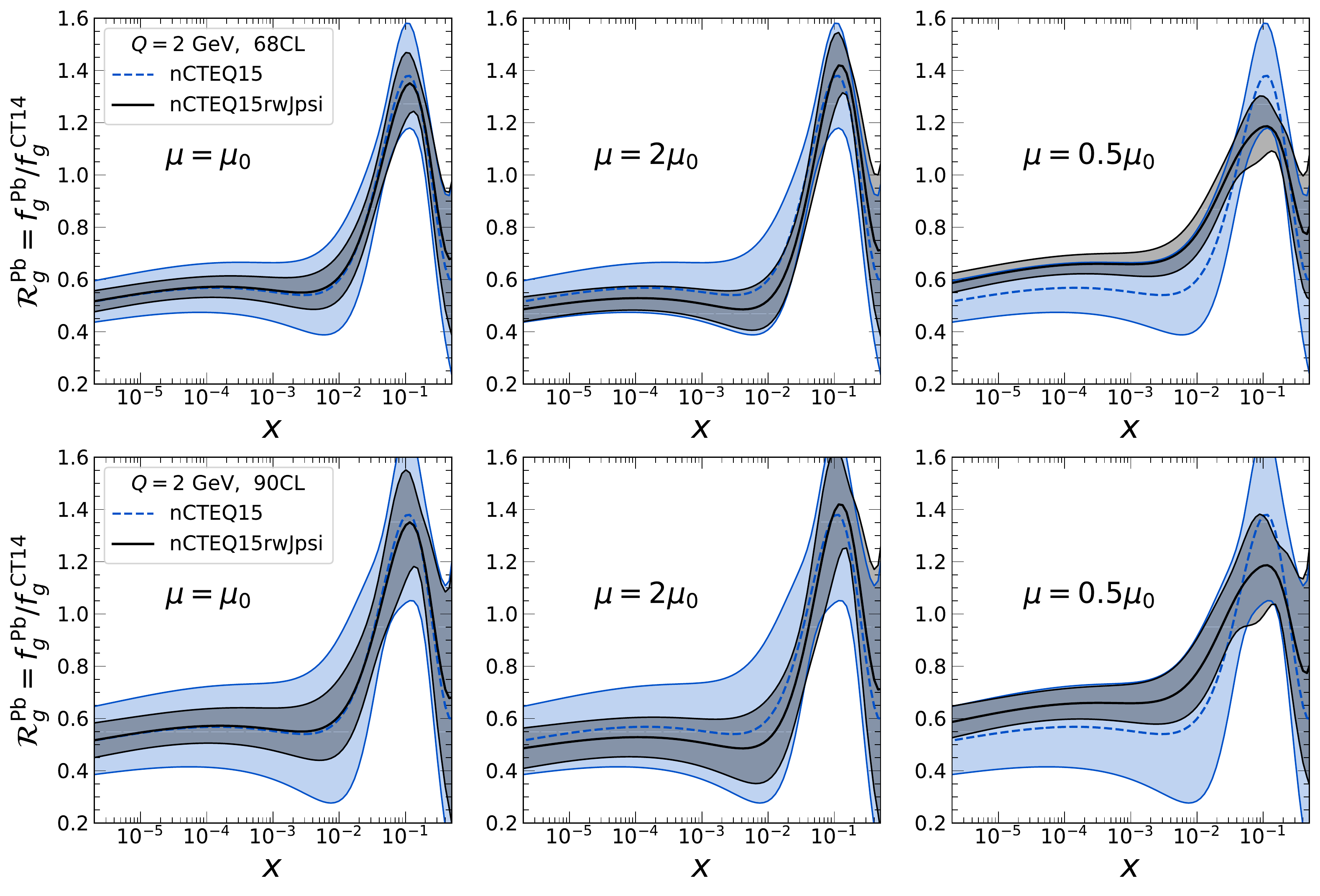}}
\\
\subfloat[EPPS16\label{fig:epps_Jpsi_68CL_vs90CL}]{
\includegraphics[width=0.90\textwidth]{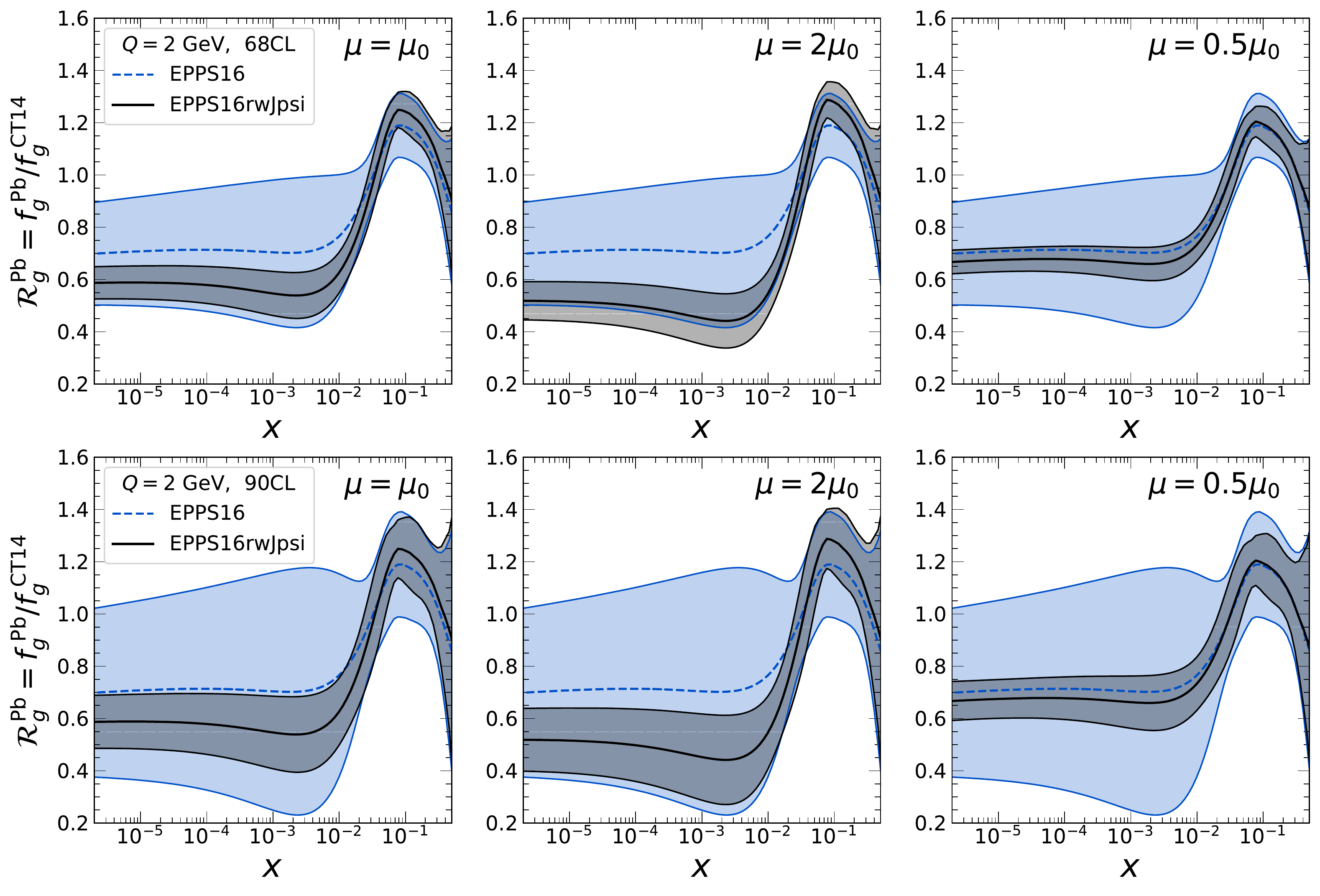}}
\caption{Gluon distribution resulting from reweighting of (a) nCTEQ15
  and (b) EPPS16 nPDFs with $J/\psi$ data. The upper rows show errors at 68\% CL
  for comparison with Fig.~1f in the original reweighting paper~\cite{Kusina:2017gkz}.
  The lower rows show the same distributions with errors at 90\% CL.}
\label{fig:Jpsi_68CL_vs90CL}
\end{figure*}

\begin{figure*}[!htb]
\centering{}
\subfloat[$D$-meson nCTEQ15\label{fig:COMBncteqD}]{
\includegraphics[width=0.95\textwidth]{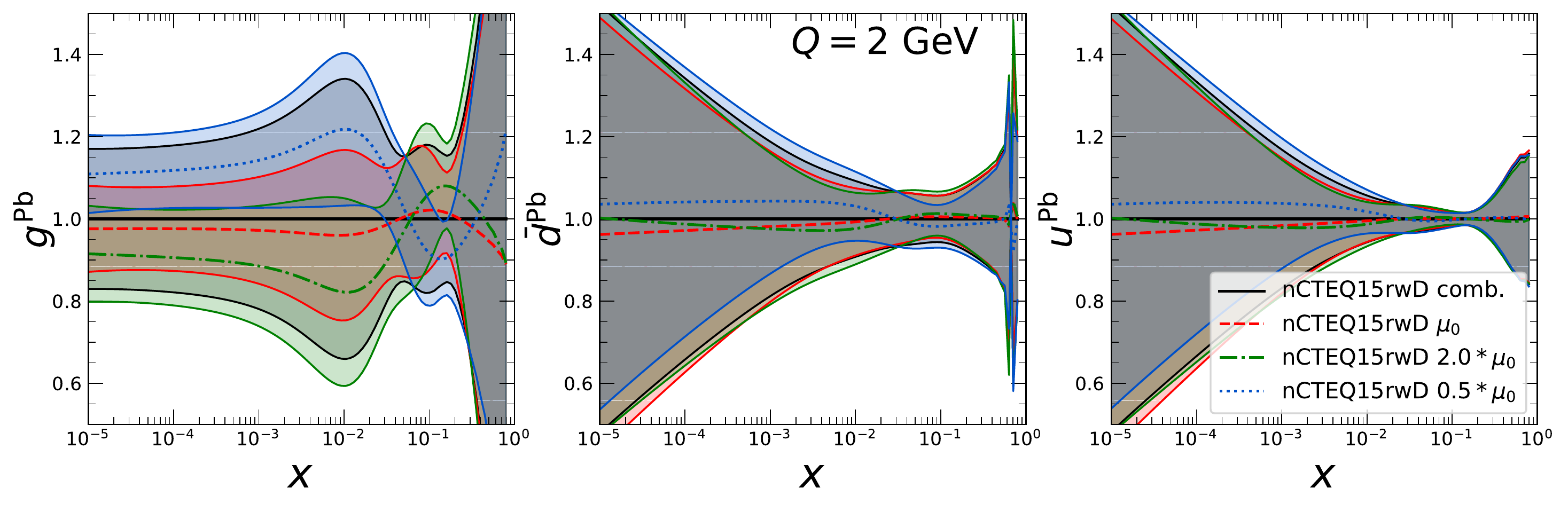}}
\\
\subfloat[$B\to J/\psi$-meson nCTEQ15\label{fig:COMBncteqB}]{
\includegraphics[width=0.95\textwidth]{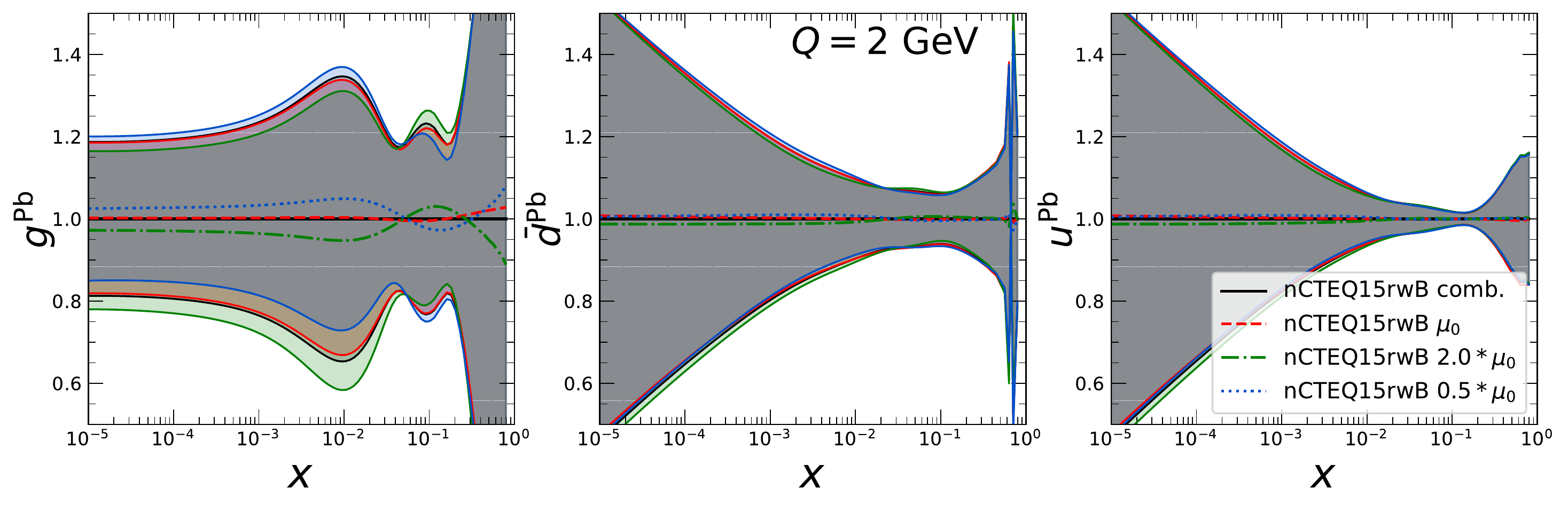}}
\\
\subfloat[$J/\psi$-meson nCTEQ15\label{fig:COMBncteqJpsi}]{
\includegraphics[width=0.95\textwidth]{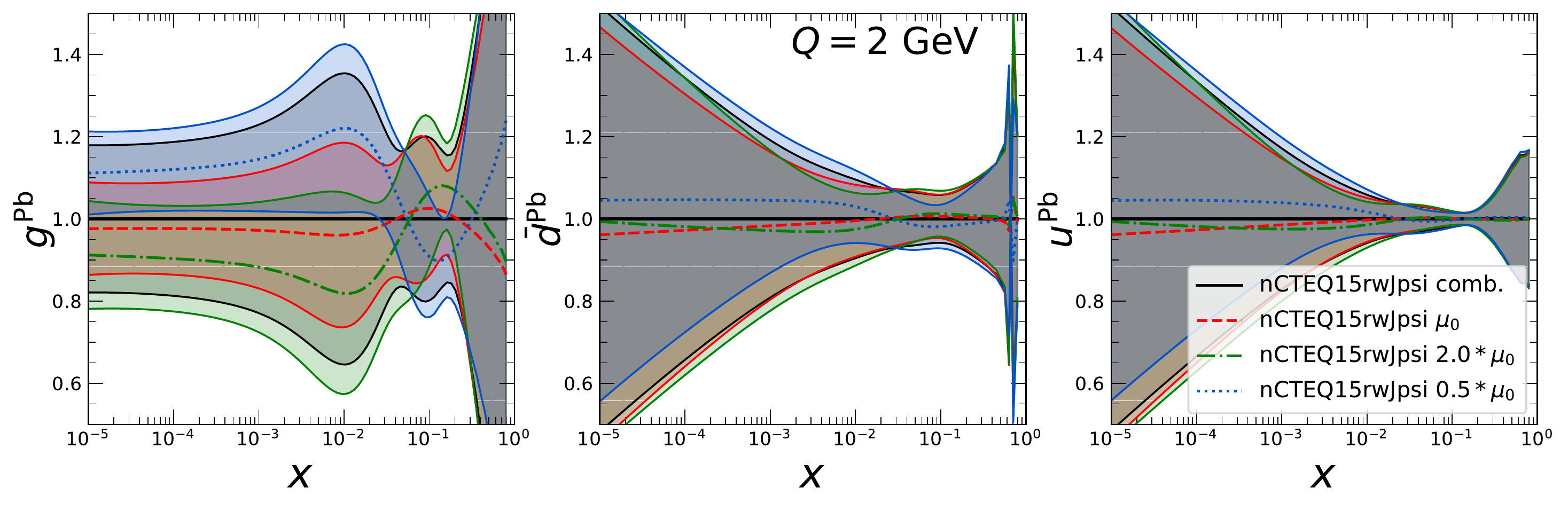}}

\caption{Comparison of the reweighting results for nCTEQ15 nPDFs with
  (a) $D$-meson,
  (b) $B\to J/\psi$, and
  (c) $J/\psi$
  data using different choice of factorization/renormalization scales.
  Additionally we show a combined set of PDFs for uncertainties from
  different scales choices were combined (solid black line).}
\label{fig:COMBncteq}
\end{figure*}

\begin{figure*}[!htb]
\centering{}
\subfloat[$D$-meson EPPS16\label{fig:COMBeppsD}]{
\includegraphics[width=0.95\textwidth]{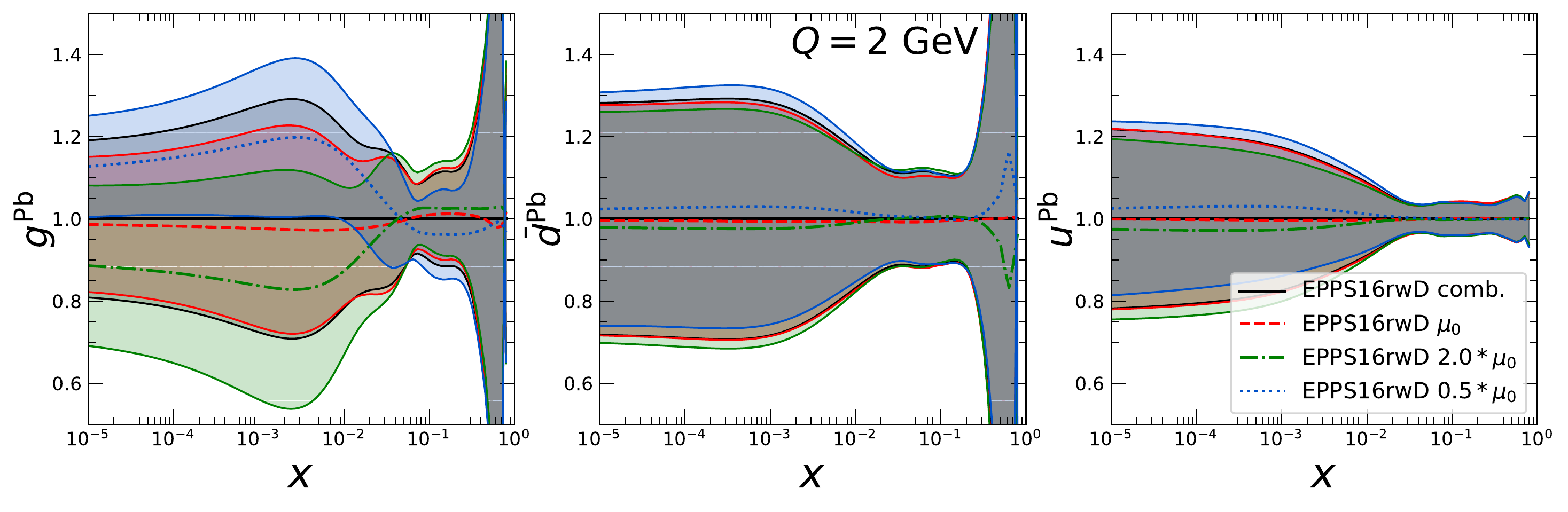}}
\\
\subfloat[$B\to J/\psi$-meson EPPS16\label{fig:COMBeppsB}]{
\includegraphics[width=0.95\textwidth]{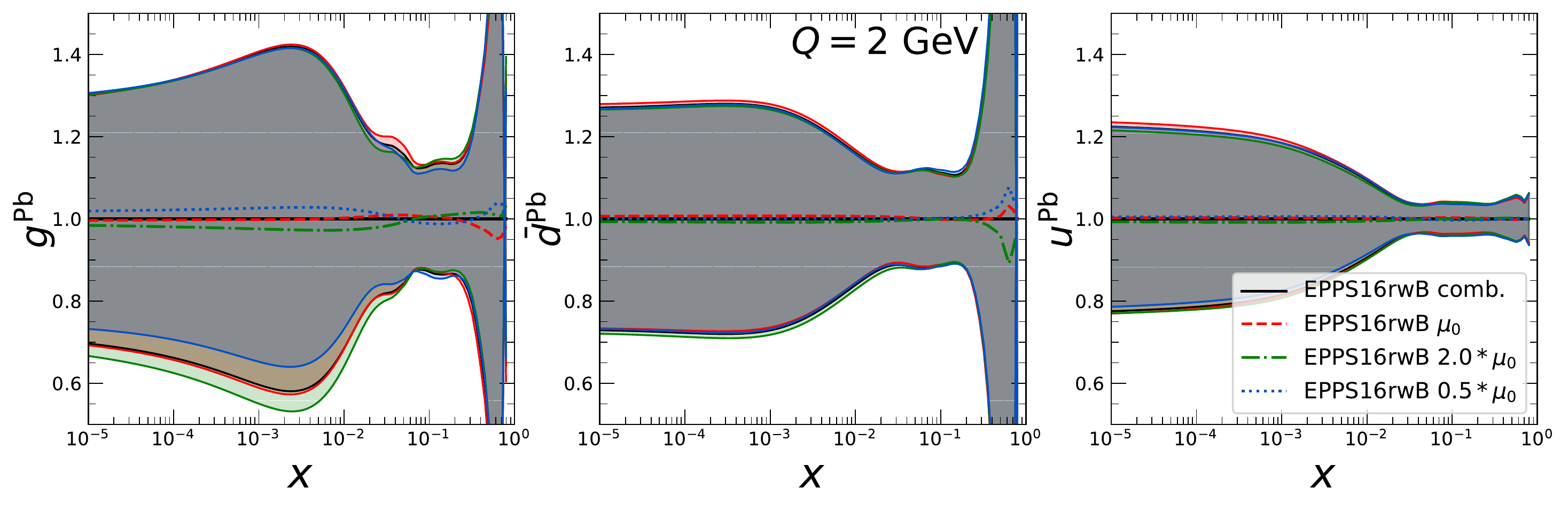}}
\\
\subfloat[$J/\psi$-meson EPPS16\label{fig:COMBeppsJpsi}]{
\includegraphics[width=0.95\textwidth]{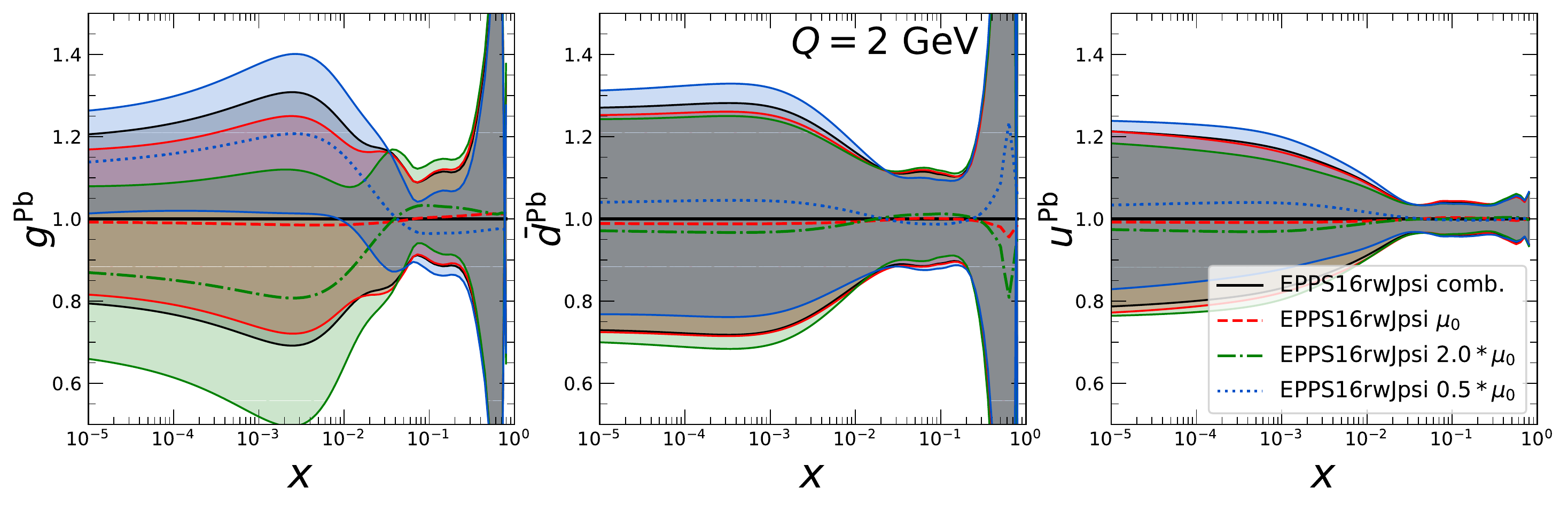}}
\caption{Comparison of the reweighting results for EPPS16 nPDFs with
  (a) $D$-meson,
  (b) $B\to J/\psi$, and
  (c) $J/\psi$
  data using different choice of factorization/renormalization scales.
  Additionally we show a combined set of PDFs for uncertainties from
  different scales choices were combined (solid black line).}
\label{fig:COMBepps}
\end{figure*}

In this appendix we provide a selection of additional plots showing the
HF RnPDFs that give additional details on the obtained results
and can be also compared with the results obtained in ref.~\cite{Kusina:2017gkz}.
We show results for the nCTEQ15 and EPPS16 $D$, $B\to J/\psi$, and $J/\psi$ RnPDFs. We restrict from showing
the results with $\Upsilon(1S)$ data as (due to the large uncertainties
of the data) the impact on the original nPDFs was very limited,
see ref.~\cite{Kusina:2017gkz}.
In Figs.~\ref{fig:D_68CL_vs90CL}, \ref{fig:B_68CL_vs90CL}, and~\ref{fig:Jpsi_68CL_vs90CL}
we present gluon NMF obtained from the reweightings with $D$, $B\to J/\psi$,
and $J/\psi$ meson data correspondingly. Figure (a) always corresponds to the
reweighting in case of nCTEQ15 nPDFs and Fig. (b) to the reweighting with the
EPPS16 nPDFs. The upper rows of Figs. (a) and (b) always show the
68\% CL results (that can be directly compared to the figures presented
in ref.~\cite{Kusina:2017gkz}), the lower rows provide the same results
but with PDF uncertainties calculated at 90\% CL.
In Figs.~\ref{fig:COMBncteq} and ~\ref{fig:COMBepps} we present results for gluon,
up quark, and anti-down quark distributions (the other distributions exhibit analogical
features). In these plots we displayed distributions obtained from the reweighting
with different scale choice \{$\mu_0$, $2\mu_0$, $0.5\mu_0$\} and a
distribution where scale uncertainties were combined. For better visibility
all PDFs were scaled by the central value of the combined distribution.
As expected the main impact is on the gluon distribution, the quark PDFs
are mostly unchanged after including the HF data.
We can also see that the uncertainty of the set with the combined scale
uncertainty is smaller than the envelope of the error bands provided by
the results for individual scale choices but larger than the uncertainty
of the central scale choice.
Generally whenever possible we recommend to use PDF sets obtained with specific
scale choice (such that it is correlated with the scale used in the theoretical
calculation). For observable were there is no reason to correlate the scales
one can either take the envelope of the three PDF sets or use the combined
PDF set to restrict the number of necessary evaluations.

\clearpage